%% file: main.tex
\documentclass[final]{IEEEtran}
\usepackage{amsthm,amssymb,graphicx,multirow,amsmath,color,amsfonts,physics}
\usepackage[update,prepend]{epstopdf}
\usepackage[noadjust]{cite}
\usepackage{tikz}
\usepackage{bbm} 
\usepackage{pdfpages}
\usepackage{balance}
\usepackage{multirow}
\usepackage{comment}
\usepackage{subfigure}

\newtheorem{corollary}{Corollary}
\allowdisplaybreaks 

\begin{document}
\include{notation}
\pagenumbering{gobble}
\graphicspath{{./Figures/}}
\title{
Performance Analysis of Infrastructure Sharing Techniques in Cellular Networks:\\ A Percolation Theory Approach}
\author{
 Hao Lin,  Mustafa A. Kishk and Mohamed-Slim Alouini
\thanks{Hao Lin is with the Electrical and Computer Engineering Program, Computer, Electrical, and Mathematical Sciences and Engineering Division (CEMSE), King Abdullah University of Science and Technology (KAUST),
Thuwal 23955-6900, Saudi Arabia (e-mail: hao.lin.std@gmail.com).\\
\indent Mustafa A. Kishk is with the Department of Electronic Engineering,
Maynooth University, Maynooth, W23 F2H6 Ireland (e-mail:
mustafa.kishk@mu.ie).\\
\indent Mohamed-Slim Alouini is with the CEMSE Division, King Abdullah
University of Science and Technology (KAUST), Thuwal 23955-6900,
Saudi Arabia (e-mail: slim.alouini@kaust.edu.sa).}
}

\maketitle
\vspace{-2cm}
\begin{abstract}
In the context of 5G, infrastructure sharing has been identified as a potential solution to reduce the investment costs of cellular networks. In particular, it can help low-income regions build 5G networks more affordably and further bridge the digital divide. There are two main kinds of infrastructure sharing: passive sharing (\ie site sharing) and active sharing (\ie access sharing), which require mobile network operators (MNOs) to share their non-electronic elements or electronic elements, respectively. Because co-construction and sharing can achieve broader coverage with lower investment, through percolation theory, we investigate how different sharing strategies can deliver large-scale continuous services. First, we examine the percolation characteristics in signal-to-interference-plus-noise ratio (SINR) coverage graphs and the necessary conditions for percolation. Second, we propose an `average coverage radius' to approximate the SINR graph with a low base station (BS) density based on the Gilbert disk model. Finally, we estimate the critical conditions of BS densities of MNOs for different sharing strategies and compare the percolation probabilities under different infrastructure sharing strategies.
\end{abstract}

\section{Introduction} \label{sec:Intro}
\indent The advent of the Fifth Generation (5G) mobile communication technology enables digital transformation and accelerates future digital economic growth. Due to its significant commercial potential, accelerating 5G network deployment has become a key priority for global mobile network operators (MNOs). At the same time, reducing network construction and operation costs, particularly in regions without advanced 5G infrastructure, has emerged as a critical challenge \cite{GSMA5GWhitepaper,ISoverview}. In \cite{8951153}, the potential of infrastructure sharing, data sharing and spectrum sharing was investigated, showing that leveraging existing public infrastructure can reduce the anticipated cost by approximately 40\% to 60\%.\\
\indent Infrastructure sharing involves sharing existing infrastructure and jointly deploying new facilities among MNOs. It has become an important research topic in recent years and a commercial reality in the 5G context \cite{cano2020evolution}. Based on which network elements that MNOs share, there are two main types of infrastructure sharing: \textit{passive sharing} and \textit{active sharing}. Passive sharing, which is also referred to as site sharing, implies the sharing of non-electrical elements at a site, such as shelter, cabinet, mast, power supply, management system. Since no additional spectrum resources are required, passive sharing allows MNOs to quickly expand their range of services by installing transceivers on other MNOs' base stations (BSs). Differently, active sharing involves the sharing of electronic components such as antennas, radio access networks, backhaul networks, and parts of core network. Therefore, active sharing is also called access sharing, where providers and operators offer access to others' resources to serve their own customers better \cite{IS}. In specific applications, infrastructure sharing is often accompanied by spectrum sharing, where different MNOs share their spectrum resources to obtain a wider bandwidth and higher data rate \cite{7876864}. Unlike traditional MNOs, mobile virtual network operators (MVNOs) do not have their own equipment and infrastructure. Instead, they provide services to their customers through other MNOs' infrastructure. Therefore, they have different sharing behaviors from MNOs \cite{7105671}.\\
\indent Since 2019, China Telecom and China Unicom have built the world's first, largest, and fastest 5G Standalone shared network, realizing one physical network correlated with two logical networks and multiple customized private networks. The two parties have built more than 1 million 5G shared BSs. At the same time, China Mobile has built more than 1.27 million 5G BSs, of which about 850,000 were jointly built and shared with China Broadcast Network. On May 17, 2023, telecom operators in China announced that they will jointly launch what they claim to be the world’s first commercial 5G inter-network roaming service trial \cite{5Groaming}. Infrastructure sharing is not only the main trend of future 5G development but also an important method to bridge the digital divide \cite{chaoub20216g}. Since 2015, the high cost of basic internet access in emerging markets, especially the poorest countries, has limited the access of the poor to the digital economy. Infrastructure sharing can reduce the substantial sunk costs of installation and provide extensive opportunities in other fields. For example, in the 6G era, infrastructure sharing could also provide opportunities for data sharing to help meet the considerable computational demands of multi-modal learning \cite{du2024distributed}.\\
\indent For cellular networks, it is important to better quantify the improvement in quality of service (QoS) arising from infrastructure sharing \cite{10077468}, which requires attention to the proportion and continuity of service coverage. Among various potential mathematical tools for analyzing coverage performance in cellular networks, percolation theory has its unique value. Percolation theory focus on whether giant connected components exist in a network \cite{Percolation,haenggi2012stochastic}. In cellular networks, percolation refers to the existence of large-scale continuous service areas. For example, users in mobile vehicles often require uninterrupted network access. In addition to the Doppler effect caused by the physical speed and the shadow effect caused by obstructions, users do not want to lose connection due to insufficient BS density. For the Internet of Vehicles, continuous services involve road information updates, the backup of driving data, navigation, and security. The monitoring network or the Internet of Things on the road should also be fully covered. For administrative boundaries or natural boundaries that need to be protected, it is also crucial to provide large-scale, continuous, safe, and reliable services. In this paper, we focus on the no sharing, passive sharing, and active sharing strategies between traditional MNOs with their own spectrum resource, transceivers and BSs, without considering MVNOs. We aim to analyze the ability of cellular networks with infrastructure sharing to form large-scale continuous coverage areas by studying discrete percolation and continuous percolation.
\subsection{Related Work}
 Infrastructure sharing, as a potential solution to improve network performance in multi-operator regions, becomes vital to reduce the high costs of BS deployment in the 5G era. However, few studies discussed the impact of infrastructure sharing on the service continuity of user equipment (UEs) from the perspective of percolation theory. For cellular networks, we consider SINR as the main metric to judge the continuity of BSs' coverage areas. Therefore, we divide the related work into: i) infrastructure sharing and ii) percolation theory and SINR analysis. \\
\indent \textit{Infrastructure Sharing}: As an important concept in 5G cellular networks, infrastructure sharing has been analyzed in various works in literature, which can save the cost, share the risk, boost quality of service and avoid environmental emissions \cite{dlamini2021remote}. There exist various models of infrastructure sharing. Authors in \cite{7105671} identified the characteristics of existing and future multi-operator network architectures, including mobile virtual network operators (MVNOs), trusted third parties, unique infrastructure providers, and standalone cases. A novel BS switching-off scheme was proposed to achieve significant energy and cost savings. Different sharing strategies have different advantages and also challenges, therefore they can be compared or combined in different application scenarios. 
In \cite{7343930}, authors assessed the fundamental trade-offs between spectrum and radio access infrastructure sharing. 
Authors in \cite{7562085} accurately modeled the channel propagation and antenna characterization and showed that a full spectrum and infrastructure sharing configuration can help increase user rate and bring economical advantages. 
In \cite{7500354}, authors introduced a mathematical framework to analyze multi-operator cellular networks that share spectrum licenses and infrastructure elements. Authors in \cite{8315130} proposed a mathematical framework to model a multi-operator mmWave cellular network with co-located BSs. They derived the SINR distribution and the coverage probability. 
In \cite{8594671}, authors proposed a multi-operator cooperation framework for sharing base stations among $N$ number of co-located radio access networks to improve energy efficiency. 
The proposed algorithm had great capacity in saving energy as well. They modeled locations of BSs in each network using independent Hardcore Poisson point processes (HCPPs). In infrastructure sharing, different MNOs' operation or sharing methods are also various.
In \cite{8329530}, authors modeled a single buyer MNO and multiple seller MNO infrastructure sharing system. Considering a given QoS in terms of the SINR coverage probability, they analyzed the trade-off between the transmit power of a BS and the intensity of BSs of the buyer MNO. Especially, MVNOs do not have their own infrastructure, so that their main expenses only come from renting transceivers and infrastructure \cite{7105671}. In \cite{zheng2017economic}, authors studied the behavior of MVNOs and Internet service providers, and derived the conditions for cross-carrier MVNOs to make profits and reduce costs for their users.
In a multi-operator cellular network, how to choose the best sharing solution requires reference to factors such as profit growth and performance improvement. Authors in \cite{cano2017optimal} considered regions with different areas and user amount, and discussed the optimal sharing strategy and number of active base stations under different service unit prices. Focusing on remote and rural areas, the authors in \cite{dlamini2021remote} reduced energy consumption while maintaining a quality of service comparable to that in urban areas. To better model random networks without losing accuracy and tractability, stochastic geometry has been widely used to evaluate the network performance with different infrastructure sharing strategies \cite{hmamouche2021new}.
They employed the homogeneous Poisson point process (PPP) and Gauss-Poisson process (GPP) to analyze the coverage probability and average user data rate. They extended their work in \cite{7876864}, differentiated the spectrum sharing experiencing flat or frequency-selective power fading, and considered the impact of network density imbalance between sharing MNOs. In summary, infrastructure sharing has been analyzed from many indicators, but there is still a lack of research on the existence of large-scale connected service areas.

\indent \textit{Percolation Theory and SINR analysis}: Percolation theory has been widely used to analyze the existence of large-scale multi-hop links or connected coverage areas in wireless networks \cite{haenggi2012stochastic, elsawy2023tutorial}. Authors in \cite{zhang2018robustness} studied the robustness of two spatially embedded networks that are interdependent. Focusing on the Gilbert disk model (GDM) and random Gilbert disk model (RGDM), authors in \cite{anjum2019percolation} derived the bounds of critical density regime in these two cases that are used to analyze the percolation in large-scale wireless balloon networks. They also analyzed the critical density of unmanned aerial vehicle (UAV) networks to ensure large-scale network coverage in \cite{9049663}. For sensing and monitoring applications, the path exposure problem was characterized in \cite{8794718,liu2012optimal}, where the authors find the critical density of sensors or cameras to detect moving objects over arbitrary paths through a given region. In \cite{9214384}, a novel and low-cost countermeasure against malware epidemics in large-scale wireless networks (LSWN) was proposed, which was denoted as spatial firewalls. In \cite{9214384} and \cite{9240972}, the critical density of such spatial firewalls and the percentage of secured devices were characterized using the tools of percolation probability. Authors in \cite{yemini2019simultaneous} proved the possibility of the coexistence of random primary and random secondary cognitive networks, both of which can include an unbounded connected component. Authors in \cite{wu2023connectivity} investigated the BS networks assisted by reconfigurable intelligent surfaces (RISs), and derived the lower bound of the critical density of RISs. Based on dynamic bond percolation, authors in \cite{han2024dynamic} proposed an evolution model to characterize the reliable topology evolution affected by the nodes and links states. In cellular networks, the effect of SINR on network performance and connectivity can not be ignored. Authors in \cite{jahnel2022sinr} derived a vital relationship between the interference cancellation factor, SINR threshold, and the status of percolation. In \cite{tobias2020signal}, authors showed that an SINR graph has an infinite connected component when the device density is large enough and the interferences are reduced sufficiently. They also estimated the relationship between the critical interference cancellation factor and device density. Authors in \cite{elsawy2023tutorial} summarized several classical percolation models. They investigated the critical relationship between device density, interference reduction factor and SINR threshold in the SINR graphs. However, there is still a lack of research that can provide critical conditions or approximate critical conditions for percolation in SINR graphs with infrastructure sharing. Therefore, it is necessary to further utilize SINR analysis to evaluate the network connectivity through percolation theory.

\subsection{Contributions}

Different from the existing literature on infrastructure sharing, this paper uses the percolation theory to study the possibility of continuous effective services under different infrastructure sharing strategies, where there are two considered MNOs in the entire network. The main contributions of this paper are as follows:\\
\indent \textit{A new perspective for performance comparison between different infrastructure sharing strategies.} In this paper, we compare the probabilities of forming large-scale continuous service areas under `no sharing', `active sharing', and `passive sharing' strategies. We show that percolation probability has its unique advantage in capturing the coverage and handover performance together. We compare the influence of MNOs' BS densities on the percolation probability under different sharing strategies, and show that `active sharing' can obtain not only the highest coverage performance but also the highest percolation probability. \\
\indent \textit{Evaluation method of SINR coverage.} In cellular networks, the impact of interference can not be ignore, and the signal-to-noise ratio (SNR) method overestimates the coverage of BSs. At the same time, the random distribution of BSs makes the coverage area of each BS irregular. Therefore, we propose an `average coverage radius' to analyze the BSs' SINR coverage. We verify that such an approximate evaluation method is reasonable through the comparison with simulated coverage probability. Computing the average coverage radius as a function of the BS densities, their transmission powers, and the SINR threshold, leads to tractable and insightful results. Especially from the perspective of percolation, we investigate the necessary conditions for large-scale continuous SINR coverage. 

\indent \textit{Percolation model and critical condition analysis.} In this paper, we confirm an important concept: continuous percolation of coverage areas can be analyzed using discrete percolation in hexagons whose side length is much less than coverage radius. We prove that when the coverage probability transits from less than $1/2$ to greater than $1/2$, the percolation probability also experiences the phase transition from zero to non-zero. Based on the `average coverage radius', we use the Gilbert disk model or the superposition of multiple Gilbert disk models to analyze the SINR coverage of BSs. We also study the critical condition of BS densities for the phase transition of percolation probability under different infrastructure sharing strategies. 

\section{System Model} \label{sec:SysMod}

Infrastructure sharing is a key concept in 5G cellular networks, and a common practice is to encourage two MNOs to share their core techs, shelters, or power cables. To understand different sharing strategies better, we focus on two MNOs (MNO $a$ and MNO $b$) and assume that the locations of BSs in each of these two MNOs follow independent Poisson point processes (PPPs) $\Phi_a$ and $\Phi_b$, respectively. The density of $\Phi_a$ is  $\lambda_a$ and the density of $\Phi_b$ is $\lambda_b$. We consider typical user equipment (UE) that subscribes to MNO $a$'s services. The basic concept of infrastructure sharing is shown in Fig.\ref{fig:SharingCases} and the differences between them are introduced as follows: 
\begin{figure}[ht]
    \centering
    \includegraphics[width=1\linewidth]{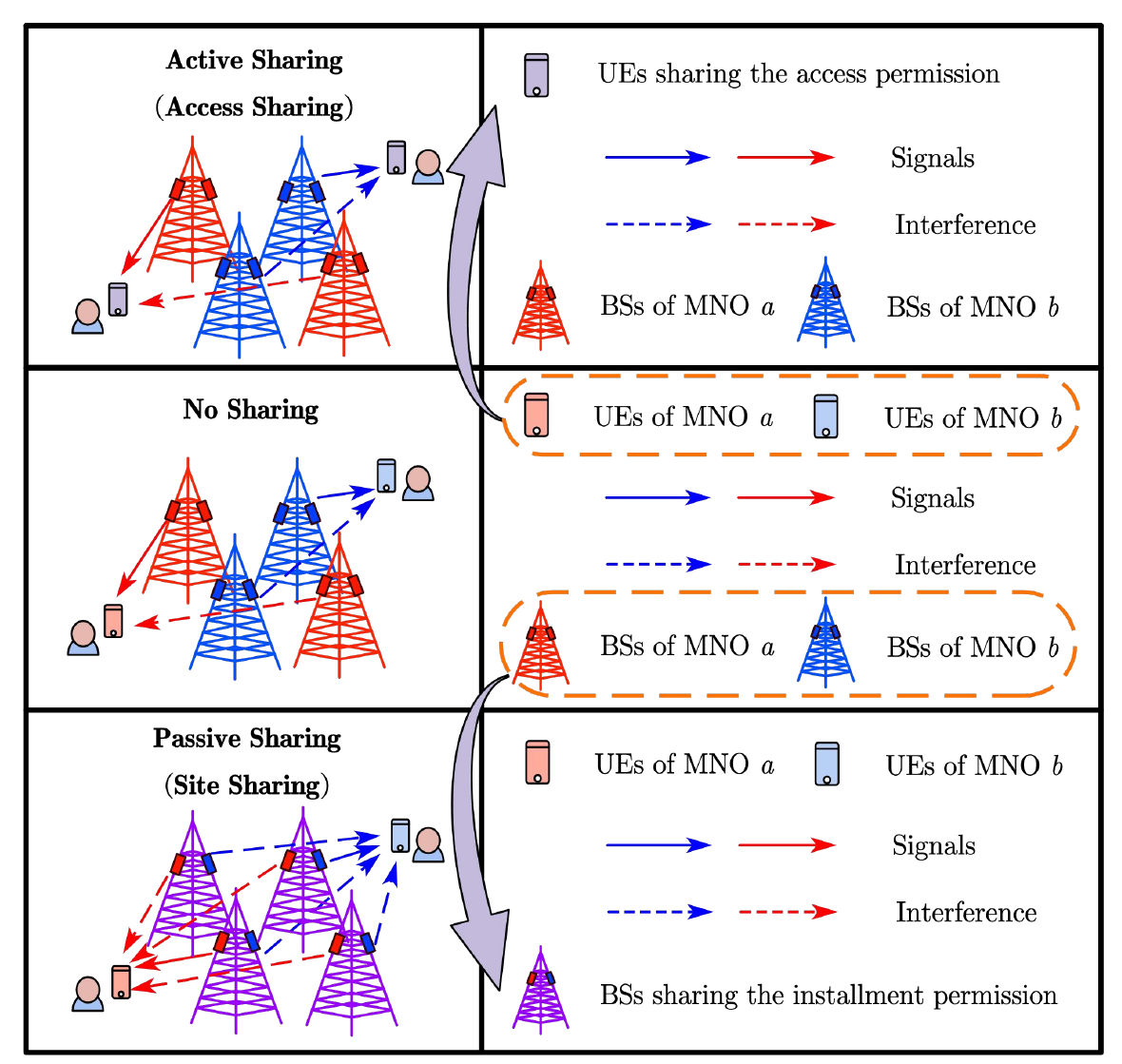}
    \caption{Comparison of `no sharing', `active sharing (access sharing)' and `passive sharing (site sharing)'. Active sharing allows UEs of each MNO to switch the spectrum and access another MNO's BSs. Passive sharing allows each MNO to deploy its transceivers on another MNO's BSs.}
    \label{fig:SharingCases}
\end{figure}

    \indent \textbf{No Sharing:} In the `no sharing' case, the UEs subscribing to services of MNO $a$ (red UEs) can only access the network through BSs that belong to MNO $a$ (the nearest red BSs). The SINR and coverage range of a BS of MNO $a$ is only affected by the signals from MNO $a$'s other BSs (other red BSs).\\
    \indent \textbf{Active Sharing:} In the `active sharing' case, MNOs share access to their networks. Therefore, UEs subscribing to MNO $a$ or MNO $b$'s service can access the Internet through all BSs. As shown in Fig.\ref{fig:SharingCases}, UEs equipped with all required antennas are marked in purple. When they access the Internet through BSs of MNO $a$ (red BSs), they are only interfered by other red BSs. When they access the Internet through BSs of MNO $b$ (blue BSs), they are only interfered by other blue BSs.\\
    \indent \textbf{Passive Sharing:} In the `passive sharing' case, MNOs share the locations of their BSs and each MNO can deploy its transceivers on other MNOs' BSs. As shown in Fig.\ref{fig:SharingCases}, BSs with transceivers from different MNOs are marked in purple, \ie purple BSs with red transceivers and blue transceivers. A typical UE subscribing to the MNO $a$'s service (red UE) can receive the signals from the red transceivers installed in the nearest purple BS. However, its service is also interfered by the red transceivers on other purple BSs. \\
\indent We consider a typical UE that subscribes to the service of MNO $a$. For the case of no sharing, it directly chooses the nearest MNO $a$'s BS to connect. For the case of active sharing, it can choose the nearest BS of MNO $a$ to connect. It can also switch the working frequency and access the nearest BS of MNO $b$. The selection depends on which MNO can provide a higher signal-to-interference-plus-noise ratio (SINR). In passive sharing, since all BSs are equipped with both MNO $a$ and MNO $b$'S transceivers, this UE can choose the nearest BS to access and communicate with the MNO $a$'s transceiver on this BS. Once the SINR at the typical UE is less than the threshold $\beta$, the typical UE can not obtain effective service from any BS. In this paper, we assume that the power from each BS received by a UE at a unit distance is $P_{a}=P_{b}=P_t$, and use $N_0$ to represent the power of noise. In general, when the signal source is located at $x_i$ and the UE is located at $z$, the SINR is expressed as:
\begin{equation}
    \beta_i(z)=\frac{P_t l(x_i-z)}{N_0+\gamma\sum\limits_{x_j\in \Phi\setminus x_i}P_t l(x_j-z)},
\label{betaiz}
\end{equation}
where $\Phi$ is the set of BSs that use the same spectrum as $x_i$ and $0\leq \gamma \leq 1$ is the interference cancellation factor. The distance attenuation $l(x)$ is written
as:
\begin{equation}
    l(x)=\left\{\begin{matrix}
 1, &\|x\|\leq 1\, {\rm m},\\
\|x\|^{-\alpha}, &\|x\|>1\,{\rm m},
\end{matrix}\right.
\end{equation}
where $\|\cdot\|$ means the Euclidean norm and $\alpha$ is the path loss exponent. \\
\indent For future communication networks, providing continuous and high-quality service for UEs is important, especially for communications in the future vision of Internet of Vehicles. This requires that the serving areas of BSs can be connected. Based on graph theory, the continuity of BSs' coverage areas can be analyzed using a random graph $G(V,E)=\{V,E\}$, where $V$ is the set of locations of BSs that provide MNO $a$'s service, and $E$ is the edge set that shows whether the coverage areas of these BSs are connected. The edge set $E$ can be expressed as:
\begin{equation}
    E=\{\overline{x_i x_j}: \exists z\in \mathbb{R}^2,\, \beta_i(z)\geq\beta\;{\rm and}\;\beta_j(z)\geq\beta\},
\label{edgeset}
\end{equation}
where $x_i,\,x_j\in V$, $\beta_i(z)$ and $\beta_j(z)$ represent the SINR from BSs located at $x_i$ and $x_j$ to $z$, respectively.\\
\indent In this paper, we focus on the coverage of downlink services. When there exist large-scale connected coverage areas of BSs, UEs moving inside such giant service areas can achieve continuous high-quality services. Based on percolation theory, we use $K\subseteq G(V,E)$ to denote a connected component inside the whole graph and let $K(0)$ denote the connected component where the origin $O$ is covered by BSs in $K$. It is worth noting that analyzing whether $K(0)$ has infinite set cardinality is equivalent to studying the probability of large-scale continuous coverage areas. Since the BS densities of two MNOs are the main factor, we define the percolation probability using $\lambda_a$ and $\lambda_b$:
\begin{equation}\label{perprodef}
    \theta(\lambda_a,\lambda_b)\overset{\triangle}{=}{\P}\{|K(0)|=\infty\},
\end{equation}
where $|K(0)|$ denotes the set cardinality of the connected component $K(0)$. In this paper, we focus on the phase transition of percolation probability from zero to non-zero, \ie the transition from $\theta(\lambda_a,\lambda_b)=0$ to $\theta(\lambda_a,\lambda_b)>0$, which is a critical indicator of the ability to form large-scale continuous service in cellular networks. For different sharing strategies, the percolation probability of the whole system $\theta(\lambda_a,\lambda_b)$ can be expressed in different forms.\\
\indent In graph theory, the Gilbert disk model (GDM) is an important tool to analyze the percolation probability in wireless networks. For a classic GDM $D(\lambda,r)$, where $\lambda$ is the BS density and $r$ is the coverage radius, the critical condition for phase transition of percolation probability is written as: \\
\begin{equation}\label{GDMcricon}
    \lambda>\frac{\lambda_c(1)}{4r^2},
\end{equation}
where $\lambda_c(1)$ is the critical density of a Gilbert disk model with the radius of $1/2$. The value of such a parameter is an open research problem where existing literature has only provided approximations and upper/lower bounds. Focusing on the continuous percolation, we estimate the value of $\lambda_c(1)$ later in this paper, particularly in Theorem \ref{theo:lambdac1}. It is worth noting that, the Gilbert disk model has been widely used in signal-to-noise ratio (SNR) graphs, where the coverage radius is not related to the BS density. In this paper, even though the edge set $E$ is defined based on SINR, we find an approximate method to theoretically analyze the critical condition of BS densities.  \\
\indent In the following section, we first investigate the properties of the SINR coverage graph and the necessary conditions for the main factors in cellular networks. We propose the `average coverage radius' to approximate the SINR coverage of each BS in low density case. Next, we give the approximate expression of critical conditions for different infrastructure sharing strategies in Sec.\ref{sec:sharing}. For ease of reading, we summarize most of the symbols in Table \ref{tab:TableOfNotations}.

\begin{table*}[htbp]\caption{Table of Notations}
\centering
\begin{center}
\resizebox{\textwidth}{!}{
\renewcommand{\arraystretch}{1}
    \begin{tabular}{ {c} | {l} }
    \hline
        \hline
    \textbf{Notation} & \textbf{Description} \\ \hline
    $\Phi_{a}$; $\lambda_a$ & The set of locations of MNO $a$'s BSs; the BS density of MNO $a$\\ \hline
    $\Phi_{b}$; $\lambda_b$ & The set of locations of MNO $b$'s BSs; the BS density of MNO $b$\\ \hline
    $G(V,E)=\{V,E\}$ & The random graph with vertice set $V$ and edge set $E$\\ \hline
    $\beta$; $\gamma$; $\alpha$ & The threshold of SINR metric; the interference cancellation factor $(0\leq \gamma \leq 1)$; the path loss exponent $(\alpha>2)$\\ \hline
    $P_t$; $N_0$ & The received power at a unit distance to a BS; the power of noise\\ \hline
    $K$; $K(0)$; $|K(0)|$ & A connected component; the connected component covering the origin; the set cardinality of $K(0)$\\ \hline
    $l(x)$; $\|x\|$ & The distance attenuation function of a distance vector $x$; the Euclidean norm of vector $x$\\ \hline
    $\Psi_i$; $\Omega_i$ & The coverage area of the BS at $x_i$; the Voronoi cell of the BS located at $x_i$\\ \hline
    $\textbf{b}(x,r)$; $D(\lambda,r)$ & The circular area centered at $x$ with radius $r$; the GDM with density $\lambda$ and coverage radius $r$\\ \hline
    $\theta(\lambda_a,\lambda_b)$ & The general expression of percolation probability of $G(V,E)$ in the whole paper\\ \hline
    $\theta(p_{cov})$ & The special expression of percolation probability which depends on the homogeneous coverage probability $p_{cov}$\\ \hline
    $\theta(\lambda,r)$ & The special expression of percolation probability of a single-layer GDM $D(\lambda,r)$\\ \hline
    $r_{a}$; $r_{b}$ & The average coverage radii of MNO $a$ and MNO $b$ in `no sharing' and `active sharing' strategies\\ \hline
    $r_{ab}$ & The average coverage radius of MNO $a$ in `passive sharing' strategy\\ \hline
    $\lambda_c(1)$ & The critical density for phase transition of percolation probability in a Gilbert disk model with half of unit coverage radius.\\ \hline
     \hline
    \end{tabular}
    }
\end{center}
\label{tab:TableOfNotations}
\end{table*}

\section{Percolation in SINR coverage graph} \label{sec:percolation}
Percolation in the SINR graph, especially regarding communications between BSs or devices, has been investigated in the related literature. However, in this paper, the SINR coverage graph is defined based on the SINR at users, which has not been investigated sufficiently before. Therefore, in this section, we introduce the properties of percolation in a `single layer SINR coverage graph', where all BSs interfere with each other:
\begin{itemize}
    \item We first define the serving areas of BSs, containing the `strongest coverage areas' (SCAs) and `coverage areas' (CAs) of BSs. We show that both of them can be used to analyze the existence of large-scale continuous serving areas, \ie percolation of BSs' coverage.
    \item Next, we introduce sub-critical cases where the SCAs of BSs can not be connected so that there is no percolation in the SINR graph. 
    \item Finally, we propose the `average coverage radius' to approximate the serving areas of BSs. We also introduce the relationship between coverage probability and phase transition of percolation probability from zero to non-zero.
\end{itemize}

\subsection{Serving areas of BSs}
\indent In this paper, we focus on the continuity of BSs' coverage areas, which depends on the value of SINR at UEs. For a typical BS at $x_i$, its coverage area (CA) can be defined as:
\begin{equation}
\begin{array}{r@{}l}
    \Psi_i
    &\overset{\triangle}{=}\bigg\{z\in\mathbb{R}^2:\beta_i(z)\geq\beta\bigg\},
\label{CA}
\end{array}
\end{equation}
where the expression of $\beta_i(z)$ is shown in (\ref{betaiz}). Especially, when the interference cancellation factor $\gamma=0$, the CA $\Psi_i$ is a circular area $\textbf{b}\big(x_i,(\frac{\beta}{P_t/N_0})^{-\frac{1}{\alpha}}\big)$ centered at $x_i$ with radius $(\frac{\beta}{P_t/N_0})^{-\frac{1}{\alpha}}$. We assume that the BSs' coverage areas always contain the circular areas around them with unit radius, therefore, ${P_t}/{N_0}>\beta$ is always satisfied. When $0<\gamma\leq 1$, the shape of CA $\Psi_i$ becomes irregular and not tractable.\\
\indent In cellular networks, it's common for users to choose the nearest BSs to them as the serving ones, which can provide the highest average signal power and best average QoS. Therefore, each BS's actual service range is contained by its Voronoi cell. Notice that the SINR of the closest BS is larger than others, we can define the Voronoi cell of the BS located at $x_i$ as:
\begin{equation}
\begin{array}{r@{}l}
    \Omega_i&\overset{\triangle}{=}\bigg\{z\in\mathbb{R}^2:\|x_i-z\|\leq \|x_j-z\|,\forall j\neq i\bigg\}\\
    &=\bigg\{z\in\mathbb{R}^2:\beta_i(z)\geq\beta_j(z),\forall j\neq i\bigg\}.
\end{array}
\end{equation}

Considering both the SINR and closest BS selection strategy, we define the `strongest coverage area' (SCA) of BS located at $x_i$ as the intersection between its CA $\Psi_i$ and its Voronoi cell $\Omega_i$, that is $\Psi_i\cap\Omega_i$, where
\begin{equation}
\begin{array}{r@{}l}
\displaystyle\Psi_i&\bigcap\Omega_i\\&\overset{\triangle}{=}\bigg\{z\in\mathbb{R}^2:\beta_i(z)\geq\beta,\;\beta_i(z)\geq\beta_j(z),\forall j\neq i \bigg\}.
\end{array}
\end{equation}
For the users inside $\Psi_i \cap \Omega_i$, they can obtain the best Internet service through the BS located at $x_i$.\\
\indent In this paper, we aim to discuss the percolation of BSs' coverage areas (CAs). To analyze percolation in SINR coverage graphs, we first introduce Theorem \ref{theo:SCAandCA} to show the relationship between the percolation of SCAs and CAs of BSs.
\begin{theorem}
    For a large-scale cellular network, the union of the strongest coverage areas (SCAs) of BSs is the same as the union of coverage areas (CAs) of BSs, that is:
\begin{equation}\label{CASCA}
\bigcup\limits_{i}\Psi_i\cap\Omega_i=\bigcup\limits_{i}\Psi_i.
\end{equation}
Therefore, the face percolation of CAs $\{\Psi_i\}$ is equivalent to the percolation of SCAs $\{\Psi_i\cap\Omega_i\}$.
\label{theo:SCAandCA}
\end{theorem}
\begin{IEEEproof}
    See Appendix~\ref{app:SCAandCA}.
\end{IEEEproof}

Therefore, for an SINR coverage graph, both of SCAs and CAs can be used to define the edges in $G(V,E)$: 
\begin{itemize}
    \item If $\Psi_i\cap\Psi_j$ is not empty, we can say that these two CA $\Psi_i$ and $\Psi_j$ are connected, \ie $\overline{x_i x_j}\in E$. 
    \item If any user on the Voronoi boundary between $\Omega_i$ and $\Omega_j$ can be covered by $\Psi_i$ and $\Psi_j$, we can say that these two strongest coverage areas $\Psi_i\cap\Omega_i$ and $\Psi_j\cap\Omega_j$ are connected, \ie $\overline{x_i x_j}\in E$. 
\end{itemize}
Especially, when there is only one BS in a large area, the required large-scale continuous coverage path can not be formed. If two adjacent BS's CAs are possible to be connected considering the effect of interference, a dense cellular network can be well-designed and almost all areas can be successfully covered. Therefore, when the BS density is at a certain value, random BS deployment can achieve large-scale continuous coverage with a non-zero probability, that is the probability of face percolation. The face percolation on the SINR coverage graph is equivalent to percolation of the random geometric graph (RGG) $G(V,E)$ where $V$ contains the location of BSs and $E$ represent the connections between BSs' CAs or SCAs. Next, using the definition of the edges through SCAs and CAs, we introduce the sub-critical cases where there is no percolation on the SINR graph.

\subsection{Sub-critical regions}
To discuss the percolation between serving areas of BSs, we start with the percolation of SCAs. A key problem is whether the points on the Voronoi boundary of neighbour BSs can be covered. If all points on Voronoi bounds can not be covered, there is no percolation on the SINR coverage graph. Therefore,  we introduce a necessary condition for the percolation of SCAs on the SINR coverage graph in Theorem \ref{theo:betagamma}.

\begin{theorem}
\label{theo:betagamma}
    For a cellular network with interference cancellation factor $\gamma$ and SINR threshold $\beta$, a necessary condition for percolation is:
\begin{equation}
    \beta\gamma<1,
\label{betagamma1}
\end{equation}
where $\beta>0$ and $0\leq \gamma\leq 1$.
\end{theorem}
\begin{IEEEproof}
    See Appendix~\ref{app:betagamma}.
\end{IEEEproof}
\begin{remark}
    It is worth noting that, when $\gamma=1$, to achieve a non-zero percolation probability, $\beta$ has to be strictly less than 1. When $0<\gamma<1$, the upper bound of $\beta$ can be higher than 1. When $\gamma=0$, this necessary condition is always satisfied, where the SINR graph becomes an SNR graph. This means both signal processing and interference controlling are significant in generating large-scale continuous services.
\end{remark}

The necessary condition (\ref{betagamma1}) can be also explained using CAs, that is, the $k$th coverage problem that we introduce in Theorem \ref{theo:devicedegree}.
\begin{theorem}
\label{theo:devicedegree}
    Consider a cellular network with interference cancellation factor $\gamma$ and SINR threshold $\beta$. Let $N$ denote `the number of potential serving BSs except for the closest one', it satisfies:
\begin{equation}
    N<\frac{1}{\beta\gamma}.
\end{equation}
\end{theorem}
\begin{IEEEproof}
    See Appendix~\ref{app:devicedegree}.
\end{IEEEproof}
\begin{remark}
    If $\beta\gamma\geq 1$, $0<\frac{1}{\beta\gamma}\leq 1$. Therefore, $N=0$. This represents that the typical user can not access the Internet service through the BSs which is not the closest one.
\end{remark}
In conclusion, we have:
\begin{itemize}
    \item When $\beta\gamma\geq 1$, $\forall \lambda>0$, any two BSs' coverage areas can not be connected. Thus, the percolation probability is 0.
    \item When $\gamma=0$, $\beta\gamma=0$ and the SINR graph reduces to an SNR graph, which can be modelled using a classic Gilbert disk model (GDM) $D(\lambda,(\frac{\beta}{P_t/N_0})^{-\frac{1}{\alpha}})$. For a classical GDM, there exists a phase transition of percolation probability from zero to non-zero with the increase in BS density $\lambda$. The critical condition for phase transition of percolation probability in a GDM is shown in (\ref{GDMcricon}).
    \item When $0<\beta\gamma<1$, whether any two BSs' SCA can be connected depends on the concrete deployment, which is not tractable. However, for the whole cellular network, the percolation probability depends on $\beta$, $\gamma$ and $\lambda$.
\end{itemize}

In this paper, we mainly discuss the case where $\beta>0$, $0<\gamma\leq1$ and $\beta\gamma<1$. Because the SINR coverage graph $G_{\rm SINR}$ is the subset of the SNR coverage graph $G_{\rm SNR}$, we introduce a necessary condition of BS density for non-zero percolation probability:
\begin{theorem}
    For a cellular network where BSs follow a PPP with density $\lambda$, a necessary condition of BS density for non-zero percolation probability is:
\begin{equation}
    \lambda >\frac{\lambda_c(1)}{4r_{\rm SNR}^2},
\label{necconforlambda}
\end{equation}
where $r_{\rm SNR}=(\frac{\beta}{P_t/N_0})^{-\frac{1}{\alpha}}$ is the SNR coverage radius.
\end{theorem}
\begin{IEEEproof}
    Let $G_{\rm SINR}$ denote the SINR coverage graph and $G_{\rm SNR}$ denote the corresponding SNR coverage graph, $G_{\rm SINR}\subseteq G_{\rm SNR}$ always holds. Therefore, no percolation on $G_{\rm SNR}$ leads to no percolation on $G_{\rm SINR}$. As shown in (\ref{GDMcricon}), we can obtain the sufficient and necessary condition for non-zero percolation probability of $G_{\rm SNR}$, which is also the necessary condition for non-zero percolation probability of $G_{\rm SINR}$.
\end{IEEEproof}
Therefore, when $\gamma\beta<1$, if the BS density $\lambda$ does not satisfy the necessary condition (\ref{necconforlambda}), percolation probability is zero. However, the value of $\lambda_c(1)$ is an open question. Using discrete percolation in a hexagonal lattice, we discuss what value $\lambda_c(1)$ converges to in Theorem \ref{theo:lambdac1}.
\begin{theorem}\label{theo:lambdac1}
    For a Gilbert disk model $D(\lambda,r)$ with vertex density $\lambda$ and coverage radius $r$, the product of $\lambda$ and $(2r)^2$ is defined as $\lambda(1)$. In a large-scale network, the critical value of $\lambda(1)$ for phase transition of percolation probability is 
\begin{equation}\label{lambdac1}
    \lambda_c(1)=\frac{4\ln 2}{\pi}.
\end{equation}
\end{theorem}
\begin{IEEEproof}
    See Appendix~\ref{app:lambdac1}.
\end{IEEEproof}

In this paper, we focus on the phase transition of percolation probability from zero to non-zero in different sharing strategies, thus the BS density $\lambda$ is the main factor we consider. Therefore, we focus on low-density cases and aim to find a proper method to approximate the coverage area of each BS. 
\subsection{Average coverage radius for low-density network}
To obtain the critical condition of BS density for phase transition of percolation probability, we first focus on a single-layer network with low BS density. Since the shapes of BSs' SINR coverage areas are irregular, we define a parameter named `average coverage radius' to help approximate the coverage areas of BSs, where we consider the global interference which is related to the distance to the closest BS. The expression of average coverage radius is introduced in Theorem \ref{theo:acr} as below:
\begin{theorem}
\label{theo:acr}
    Consider a cellular network where all BSs interfere with each other. The locations of BSs follow a PPP with BS density $\lambda$ which is low enough. In this case, the `average coverage radius' of any BS in this network is the unique solution of $\mathcal{F}(r_m,\lambda)=1$, where

\begin{equation}
    \mathcal{F}(r_m,\lambda)=\frac{\beta}{P_t/N_0}r_{m}^{\alpha}+\frac{2\pi\gamma\beta\lambda}{\alpha-2}r^2_{m}.
\label{betaratio}
\end{equation}
The `average coverage radius' can be rewritten as a decreasing function of $\lambda$, i.e.
\begin{equation}
    r=r_m(\lambda).
\label{rewrite}
\end{equation}

\end{theorem}
\begin{IEEEproof}
    See Appendix~\ref{app:acr}.
\end{IEEEproof}

\indent Especially, if the path loss exponent $\alpha=4$, the `average coverage radius' can be written as a closed form expression of BS density $\lambda$, which is shown in Corollary \ref{cor:alpha4}.
\begin{corollary}\label{cor:alpha4}
    When $\alpha=4$, the average coverage radius $r_{a}$ can be written as a closed form expression:
\begin{equation}
    r_{m}(\lambda)=\sqrt{\sqrt{\frac{\beta_0}{\beta}+\bigg(\frac{\pi\gamma\lambda\beta_0}{2}\bigg)^2}-\frac{\pi\gamma\lambda\beta_0}{2}},
\end{equation}
where $\beta_0=P_t/N_0$.
\end{corollary}
\begin{IEEEproof}
    Substitute $\alpha=4$ into Theorem \ref{theo:acr}.
\end{IEEEproof}

\indent When $\gamma$ is small enough (approaches 0), the interference can be reduced efficiently, thus the SINR graph can be approximated well using the Gilbert disk model. In (\ref{betaratio}), $r_m$ depends on the product of $\gamma\lambda$. Therefore, especially for a low BS density network, $r_m$ can approximate the coverage areas well. Also, because the locations of BSs are homogeneous, it is feasible to adopt this method to analyze the critical condition for phase transition of percolation probability. \\
\indent Adopting the average coverage radius, we can approximately model a cellular network with a low BS density as a Gilbert disk model with BS density $\lambda$ and coverage radius $r_m$, \ie $D(\lambda,r_m)$. Next, we introduce some properties of percolation probability in the classical Gilbert disk model, where the coverage probability is a significant indicator.
\subsection{Continuous Percolation in Gilbert disk model}
\indent In this paper, we aim to adopt the Gilbert disk model to discuss the phase transition of percolation probability in `no sharing', `active sharing', and `passive sharing' strategies, respectively. Therefore, we need to provide the properties of continuous percolation in the Gilbert disk model. Firstly, based on a classical Gilbert disk model, the critical condition of coverage probability for phase transition of percolation probability in Theorem \ref{theo:percov}.
\begin{theorem}\label{theo:percov}
    Consider a large-scale wireless communication system where the distribution of BSs is homogeneous and all users have the same probability of being covered. If and only if the coverage probability $p_{cov}$ is larger than $1/2$, the percolation probability $\theta$ is non-zero. Such a phase transition of percolation probability can be expressed as: 
\begin{equation}\left.
    \begin{array}{@{}r@{}l}
    \theta(p_{cov})=0, &\;{\rm if}\;p_{cov}\leq1/2,\\
    \theta(p_{cov})>0, &\;{\rm if}\;p_{cov}>1/2.
    \end{array}\right.
    \label{percov}
\end{equation}
\end{theorem}
\begin{IEEEproof}
    See Appendix~\ref{app:pcov}.
\end{IEEEproof}
\indent Using this critical condition for phase transition of the probability of continuous percolation, we show the critical condition of BS densities in the next section.

\section{Percolation on Different Sharing Strategy}
\label{sec:sharing}
In Sec.\ref{sec:percolation}, we have provided the properties of a single-layer SINR coverage graph. In this section, adopting the `average coverage radius' and properties of continuous percolation under GDMs, we discuss the phase transition of percolation probability in different sharing strategies in detail. Considering the UEs subscribing to the service of MNO $a$, we simplify the coverage model under these three sharing strategies as:
\begin{itemize}
    \item \textbf{No sharing}: A single-layer GDM $D(\lambda_a,r_{a})$ with BS density $\lambda_a$ and the coverage radius $r_{a}$.
    \item \textbf{Active sharing}: A union of two independent GDMs $D(\lambda_a,r_{a})$ and $D(\lambda_b,r_{b})$. For MNO $a$, the BS density is $\lambda_a$ and the coverage radius is $r_{a}$. For MNO $b$, the BS density is $\lambda_b$ and the coverage radius is $r_{b}$.
    \item \textbf{Passive sharing}: A single-layer GDM $D(\lambda_a+\lambda_b,r_{ab})$ with BS density $\lambda_a+\lambda_b$ and the coverage radius $r_{ab}$.
\end{itemize}

\begin{figure*}[htbp]
    \centering
    \includegraphics[width=0.75\linewidth]{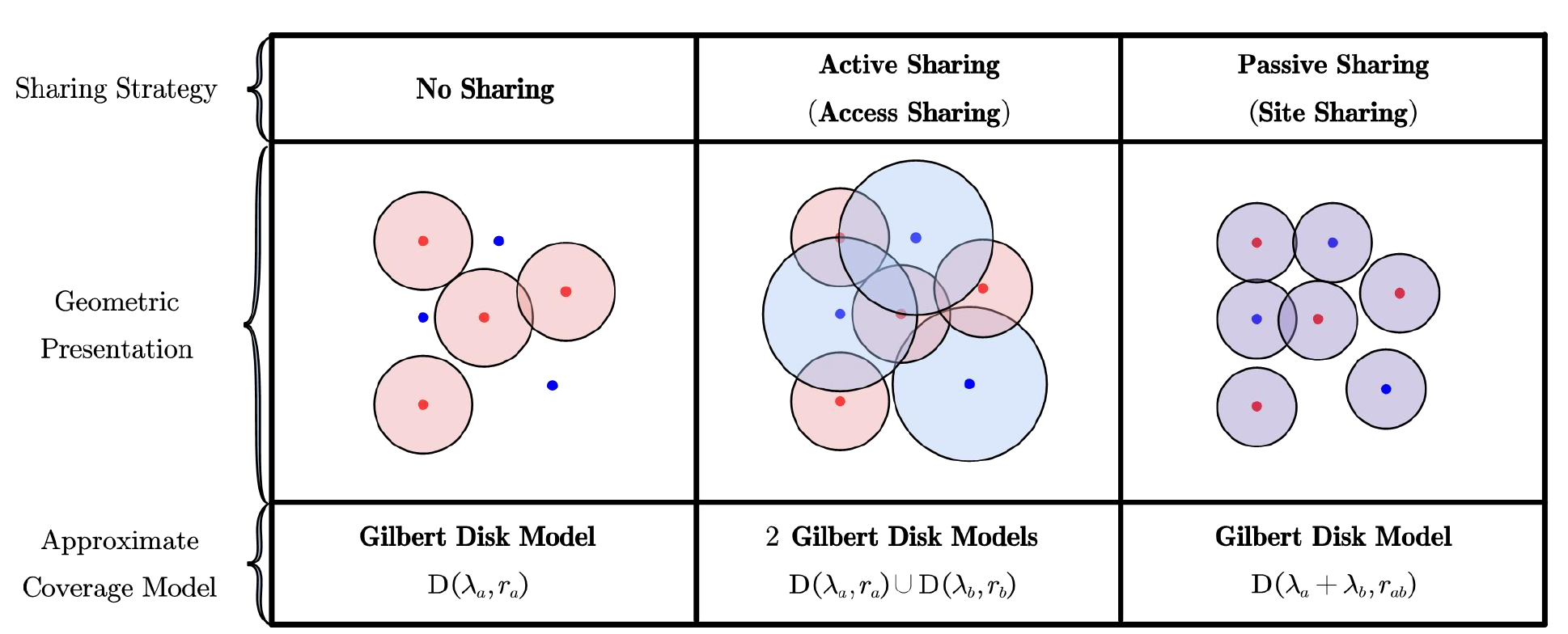}
    \caption{Mathematical models and geometric interpretations considered when studying different sharing strategies based on the Gilbert disk model.}
    \label{fig:Geometric}
\end{figure*}
\subsection{No Sharing}
\indent In the `no sharing' strategy, we consider a typical UE of MNO $a$ that can be served by MNO $a$'s BSs. The density of MNO $a$'s BSs is $\lambda_a$ and the vertice set $V=\Phi_a$. The RGG $G(V,E)$ can be approximately modelled as a GDM $D(\lambda_a,r_{a})$ and the critical condition for phase transition of percolation probability of $G(V,E)$ can be approximately expressed as:
\begin{equation}\label{ama2}
    \lambda_a r_{a}^2>\frac{\lambda_c(1)}{4}.
\end{equation}
\indent As introduced in Theorem \ref{theo:acr}, the average coverage radius is:
\begin{equation}
    r_a=r_m(\lambda_a).
\end{equation} Since $r_{a}$ depends on $\lambda_a$, we further propose the critical condition of BS density in Lemma \ref{lem:criconlam}.
\begin{lemma}\label{lem:criconlam}
    In `no sharing' strategy, the critical condition for phase transition of percolation probability can be written as the critical condition for MNO $a$'s BS density $\lambda_a$:
\begin{equation}\label{criconlama}
    \lambda_a>\frac{\lambda_c(1)}{4}\bigg(\frac{\beta_0}{\beta}\bigg(1-\frac{\pi \gamma \beta\lambda_c(1)}{2(\alpha-2)}\bigg)\bigg)^{-\frac{2}{\alpha}},
\end{equation}
where $\beta_0=P_t/N_0.$
\end{lemma}
\begin{IEEEproof}
    See Appendix~\ref{app:criconlam}.
\end{IEEEproof}
\indent In the condition (\ref{criconlama}), the critical value of $\lambda_a$ for phase transition of percolation probability in `no sharing' is related to the parameters  $\beta$, $P_t$, $N_0$, $\gamma$ and $\alpha$.

\subsection{Active Sharing}
In the `active sharing' case, MNOs share their access rights without deploying more BSs and transceivers. The vertice set $V=\Phi_a \cup \Phi_b$. The MNO $a$'s network and MNO $b$'s network do not interfere with each other. When the UE accesses MNO $a$'s network, the considered BS density is $\lambda_a$ and average coverage radius is $r_a=r_m(\lambda_a)$. Similarly, when the UE accesses MNO $b$'s network, the average coverage radius is: 
\begin{equation}
    r_{b}=r_m(\lambda_b),
\end{equation}
where $\lambda_b$ is the density of BSs of MNO $b$.

\indent Because BSs of MNO $a$ and MNO $b$ have different coverage radii $r_{a}$ and $r_{b}$ and different densities $\lambda_a$ and $\lambda_b$ respectively, we need to find the method to analyze the critical condition for percolation probability. We introduce the restriction region of the critical BS density condition of the superposition of two GDMs in Lemma \ref{lem:2GDpre}.
\begin{lemma}\label{lem:2GDpre}
    For the superposition of Gilbert disk models $D(\lambda_a,r_{a})$ and $D(\lambda_b,r_{b})$ with different densities and radii. The restriction region of the critical BS density condition is
    \begin{equation}\label{limit}
        \frac{\lambda_c(1)}{4\max\{r_{a}^2,r_{b}^2\}}<\lambda_a+\lambda_b<\frac{\lambda_c(1)}{4\min\{r_{a}^2,r_{b}^2\}},
    \end{equation}
where $r_{a}=r_{m}(\lambda_a)$ and $r_{b}=r_{m}(\lambda_b)$.
\end{lemma}
\begin{IEEEproof}
    See Appendix~\ref{app:2GDpre}.
\end{IEEEproof}

Further, using the conclusion in Theorem \ref{theo:percov}, we introduce the critical BS density condition for phase transition of percolation probability in Lemma \ref{lem:2GD}.
\begin{lemma}\label{lem:2GD}
    When there are two kinds of BSs: i) BSs of MNO $a$ with density $\lambda_a$ and coverage radius $r_{a}$ and ii) BSs of MNO $b$ with density $\lambda_b$ and coverage radius $r_{b}$ at the same time, the condition for phase transition of percolation probability of all BSs' coverage areas is written as:
\begin{equation}
    \lambda_a r_{a}^2+\lambda_b r_{b}^2>\frac{\lambda_c(1)}{4},
\end{equation}
where $r_{a}=r_{m}(\lambda_a)$ and $r_{b}=r_{m}(\lambda_b)$.
\end{lemma}
\begin{IEEEproof}
    See Appendix~{\ref{app:2GD}}.
\end{IEEEproof}
\indent We notice that the critical condition in Lemma \ref{lem:2GD} satisfies the restriction in (\ref{limit}). 

\subsection{Passive Sharing}
\indent In the `passive sharing' case, MNO $a$ deploys its transceivers on all BSs, including its own BSs and MNO $b$'s BSs, therefore, $V=\Phi_a\cup\Phi_b$. Since interference is caused by the signals from all BSs, the average coverage radius is:
\begin{equation}
    r_{ab} = r_m(\lambda_a+\lambda_b).
\end{equation}

In this case, the critical condition for phase transition of percolation probability can be approximately expressed as:
\begin{equation}\label{pscricon}
    (\lambda_a+\lambda_b)r_{ab}^2>\frac{\lambda_c(1)}{4}.
\end{equation}
\indent Since we consider that `passive sharing' only increases the density of transceivers, the expression of the average coverage radius is also similar to that of `no sharing'. Therefore, the critical condition (\ref{pscricon}) can be also written as the critical condition for MNOs' BS densities, \ie
\begin{equation}
    \lambda_a+\lambda_b>\frac{\lambda_c(1)}{4}\bigg(\frac{\beta_0}{\beta}\bigg(1-\frac{\pi \gamma \beta\lambda_c(1)}{2(\alpha-2)}\bigg)\bigg)^{-\frac{2}{\alpha}}.
\end{equation}

\section{Simulation results and discussion}\label{sec:simulation}

\begin{figure}[ht]
    \centering
    \includegraphics[width=0.9\linewidth]{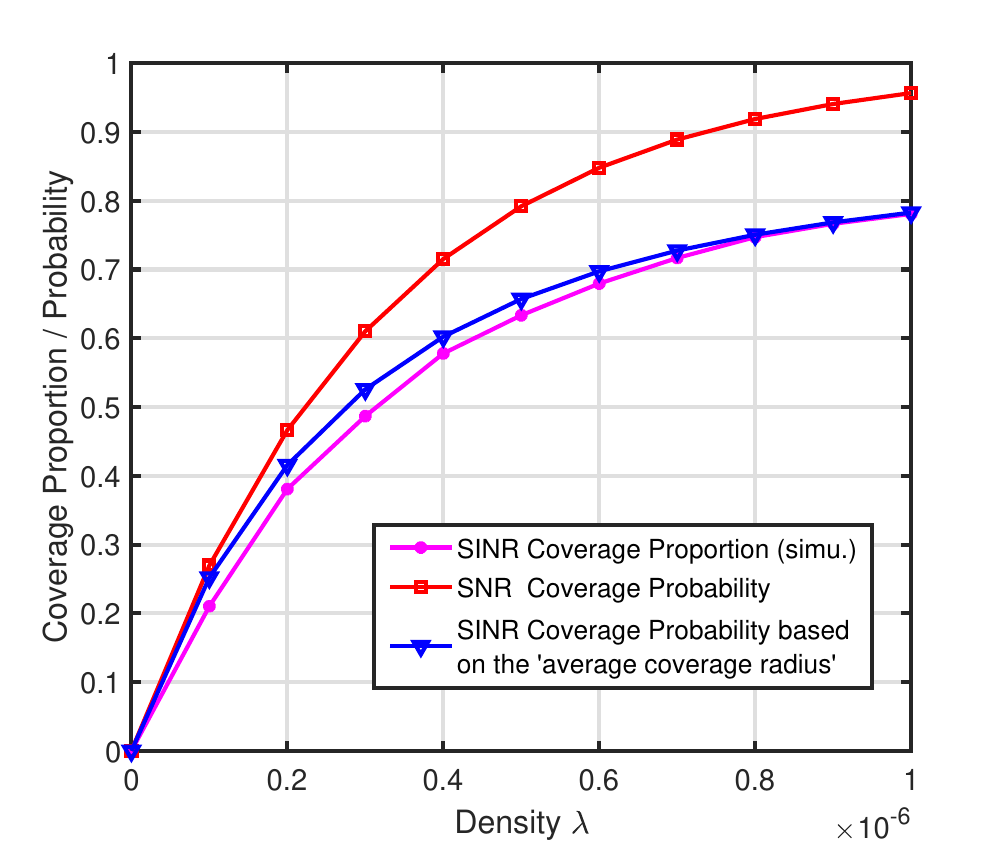}
    \caption{The relationship between the proportion of areas effectively covered by BSs providing the same MNO's service.}
    \label{fig:Covpro}
\end{figure}

\indent In this paper, a novel concept is to use the `average coverage radius' and Gilbert disk model to approximate the coverage areas of BSs. First, we prove that such a method is valid to analyze the phase transition of percolation probability. We focus on an SINR graph where there is only one MNO. Following \cite{kouzayha2022coexisting,isabona2023accurate,kumari2019short}, we set $P_t=13 {\rm\,dB}$, $N_0=-104\,{\rm dB}$ and $\alpha=4$ to simulate a cellular network in a $4000\,{\rm m}\times 4000\,{\rm m}$ residential area with micro-cell BSs. We choose $\beta=-3\,{\rm dB}$, $\gamma=1$ to satisfy the necessary conditions for percolation (\ref{betagamma1}) and conduct 100,000 Monte Carlo experiments.  
Fig.\ref{fig:Covpro} shows the SINR coverage proportion of MNO $a$ through simulation and the theoretic curve of SINR coverage probability based on the `average coverage radius'. It shows that the average coverage radius does not overestimate the actual SINR coverage probability.

\begin{figure}[htbp]
\centering
\subfigure[Percolation probability in `no sharing' strategy.]{
\begin{minipage}[t]{1\linewidth}
\centering 
\includegraphics[width=0.85\textwidth]{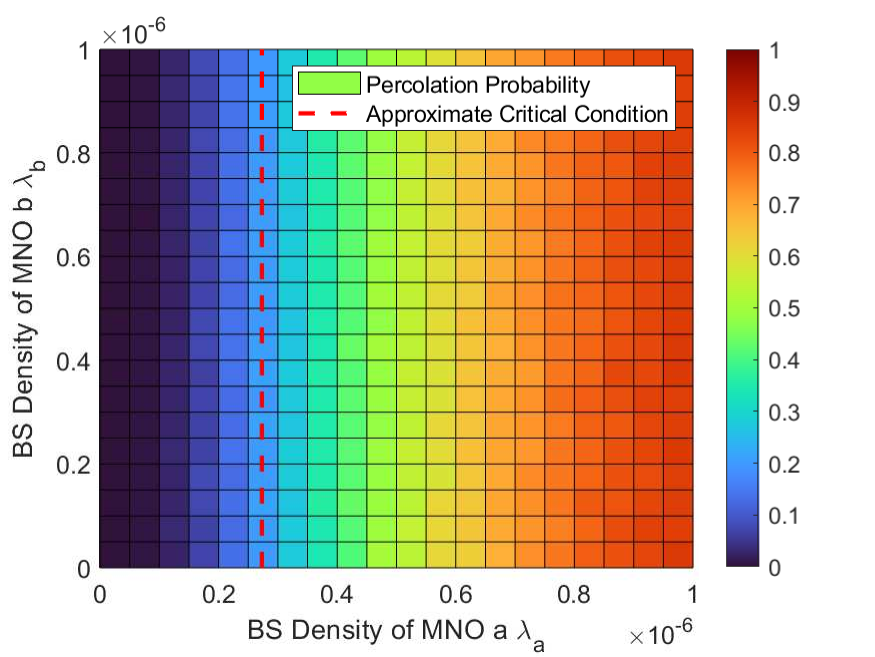}
\label{fig:PerP_ns}
\end{minipage}}

\subfigure[Percolation probability in `active sharing' strategy.]{
\begin{minipage}[t]{1\linewidth}
\centering 
\includegraphics[width=0.85\textwidth]{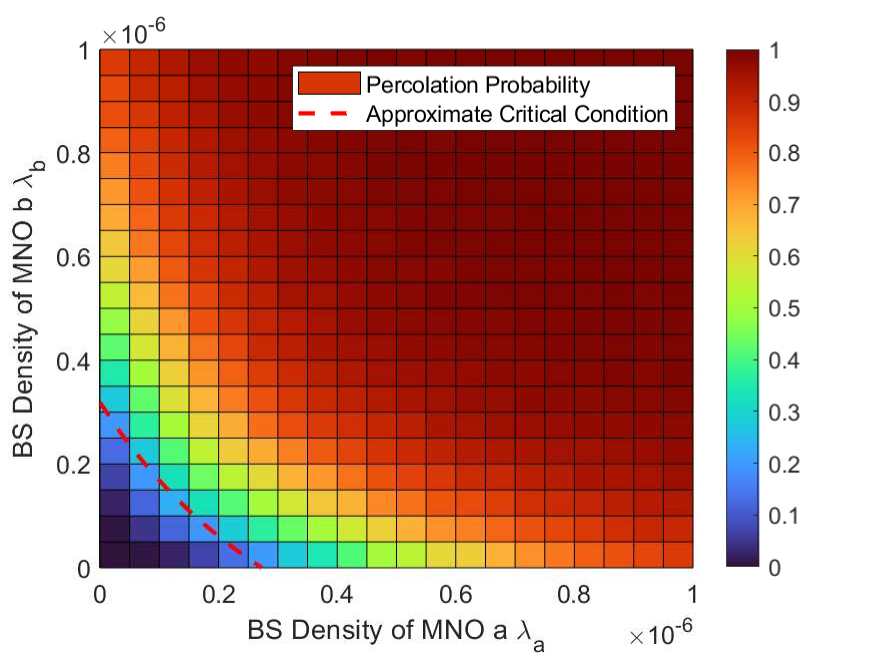}
\label{fig:PerP_as}
\end{minipage}}

\subfigure[Percolation probability in `passive sharing' strategy.]{
\begin{minipage}[t]{1\linewidth}
\centering 
\includegraphics[width=0.85\textwidth]{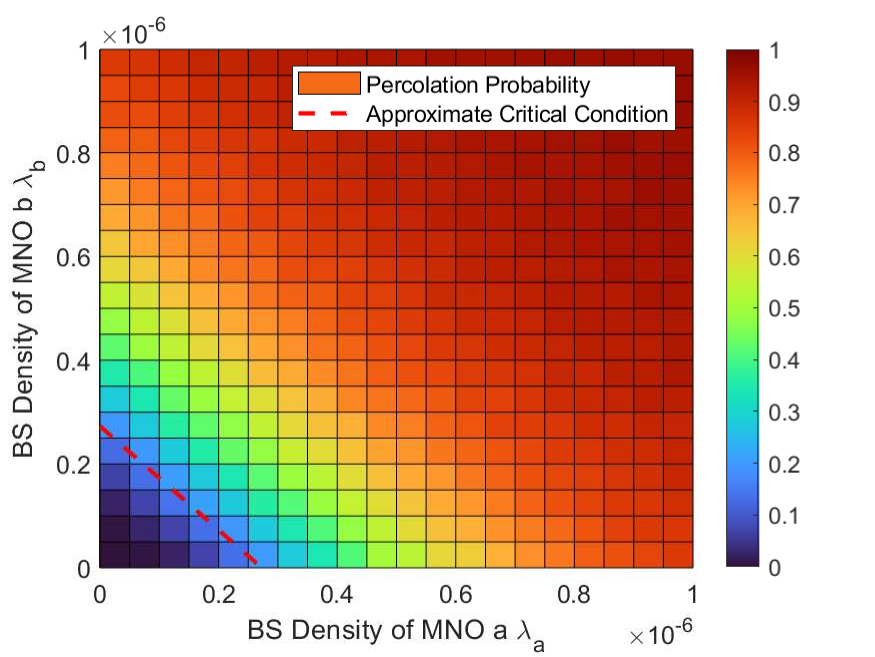}
\label{fig:PerP_ss}
\end{minipage}}

\caption{Percolation probability in different sharing strategies.}
\label{fig:PerP}
\end{figure}

\begin{figure}[htbp]
\centering
\subfigure[SINR coverage proportion in `no sharing' strategy.]{
\begin{minipage}[t]{1\linewidth}
\centering 
\includegraphics[width=0.85\textwidth]{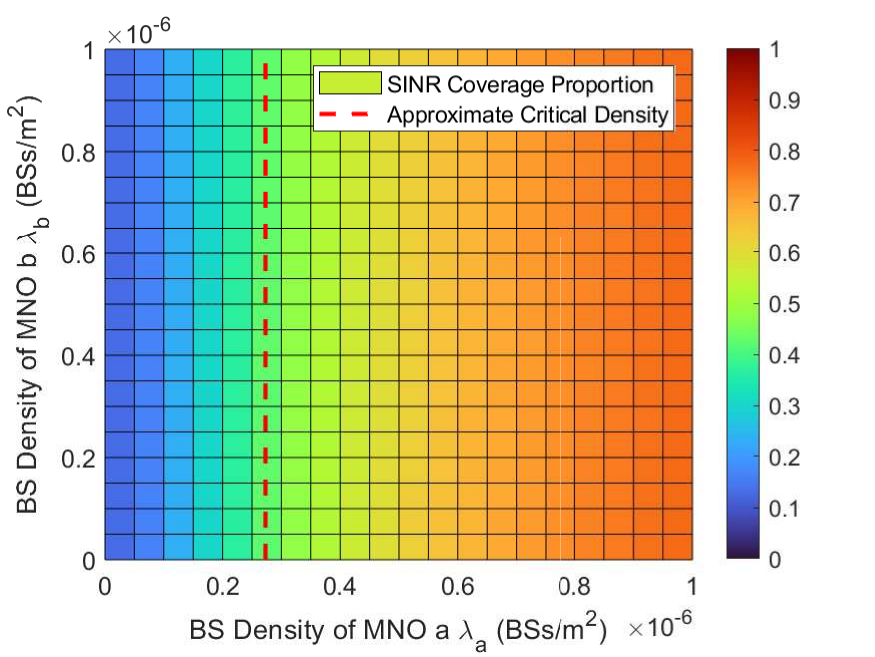}
\label{fig:Prop_ns}
\end{minipage}}

\subfigure[SINR coverage proportion in `active sharing' strategy.]{
\begin{minipage}[t]{1\linewidth}
\centering 
\includegraphics[width=0.85\textwidth]{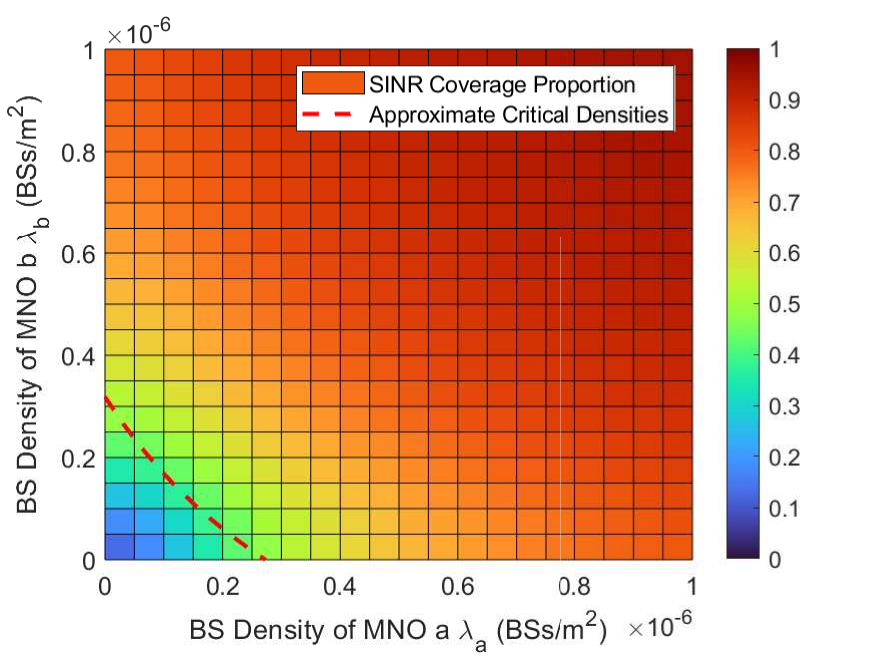}
\label{fig:Prop_as}
\end{minipage}}

\subfigure[SINR coverage proportion in `passive sharing' strategy.]{
\begin{minipage}[t]{1\linewidth}
\centering 
\includegraphics[width=0.85\textwidth]{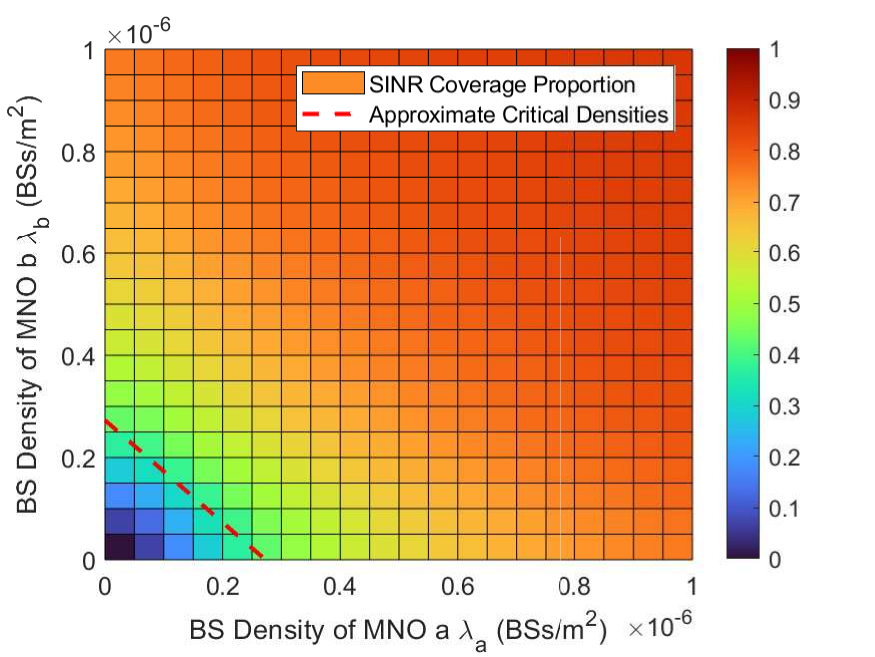}
\label{fig:Prop_ss}
\end{minipage}}

\caption{SINR coverage proportion in different sharing strategies.}
\label{fig:Prop}
\end{figure}

\indent Next, we can observe the relationship between percolation probability and SINR coverage proportion in Fig.\ref{fig:PerP} and Fig.\ref{fig:Prop}. Especially, focusing on the `no sharing' strategy, the phase transition of percolation probability happens when the BS density is within the range $(2\times 10^{-7},3\times 10^{-7})\;{\rm BSs/m^2}$, which means that approximate critical BS density (which is $2.73\times 10^{-7}\;{\rm BSs/m^2}$) obtained using our method in (12) is valid. When the density is less than the critical value, the percolation probability is less than $0.1$. Theoretically, broadening the simulation range and increasing simulation times can make the transition from a low level ($<0.1$) to a higher level be sharper. In each of the three sharing strategies, such a phase transition is accompanied by the increase in the SINR coverage proportion from $p_{cov}<1/2$ to $p_{cov}>1/2$, which is shown in Fig.\ref{fig:PerP} and Fig.\ref{fig:Prop}.

\indent From the perspective of the ability to form large-scale continuous service areas, we aim to obtain a high percolation probability. Suppose that a network with a percolation probability larger than 0.9 can be judged to have high global connectivity. Considering a network where MNO $b$'s BS density $\lambda_b=5\times10^{-7}\;{\rm BSs/m^2}$, no sharing strategy requires MNO $a$ to deploy its BSs at a density larger than $1\times10^{-6}\;{\rm BSs/m^2}$. Under active sharing strategy, the required BS density is $2.5\times10^{-7}\;{\rm BSs/m^2}$. Differently, under the passive sharing strategy, the required BS density is $5\times10^{-7}\;{\rm BSs/m^2}$, which is larger than that of the active sharing strategy but smaller than that of no sharing strategy. \\
\indent With the same BS densities, these three infrastructure sharing strategies show different percolation probability. For example, we consider the case where $\lambda_{a}=\lambda_{b}=5\times10^{-7}\;{\rm BSs/m^2}$. Without cooperation, `no sharing' strategy leads to only about 0.5 percolation probability. Under `passive sharing' strategy, the percolation probability can reach 0.85, while the `active sharing' leads to almost 1 percolation probability.\\
\indent Based on the simulation results, we can give suggestions to different MNOs. For the new MNOs with few BS deployment, our critical condition of phase transition can help them make the initial plan to realize a large-scale continuous service area at the lowest cost, which depends on the infrastructure sharing strategy that it adopts. For MNOs whose BS deployment is not enough (e.g. $\lambda<5\times10^{-7}\;{\rm BSs/m^2}$), active sharing and passive sharing can both help them further generate large-scale continuous service areas to a great extent. Compared to active sharing which performs better, passive sharing does not require the antenna and transceivers of other MNOs which might be expensive. For an MNO whose BS density is high enough (e.g. $\lambda>5\times10^{-7}\;{\rm BSs/m^2}$), cooperation with a new MNO with a low BS density is a good choice, where it can not only achieve a higher percolation probability at a low cost, but also earn profit from sharing their infrastructure.\\
\indent Consequently, this manuscript adopts percolation theory to provide connectivity analysis for cooperation between different MNOs. MNOs can analyze the connectivity improvement brought by infrastructure sharing based on their own and their partners' capabilities. In addition, MNOs need to consider the benefits that infrastructure sharing can bring. Regardless of the sharing strategy adopted, MNOs still need to consider the market changes after sharing. For two MNOs who adopt active sharing, since most UEs are already equipped with antenna for different MNOs, they need to further consider the potential risks and costs that may be brought by sharing access rights and protocols. For two MNOs who adopt passive sharing, they need to consider the cost of renting BSs or the profit of renting out BSs, as well as the installation cost and maintenance cost of more transceivers. In addition, this manuscript focuses on the percolation of SINR graphs. In the future, factors affecting service performance such as network congestion need to be considered, and dynamic percolation needs to be further discussed.

\section{Conclusion}
In this paper, we built mathematical models to analyze the connectivity of coverage areas in different infrastructure sharing strategies. We emphasized that percolation probability has its unique advantage in evaluating the ability to generate large-scale continuous coverage areas. Firstly, we analyzed the necessary conditions for percolation in SINR coverage graphs. Based on GDMs, we discussed the relationship between percolation probability and coverage probability. After that, we proposed an approximate tool to study the coverage of SINR graphs, \ie the `average coverage radius'. Using `average coverage radius', we studied the critical condition for phase transition of percolation probability under different infrastructure sharing strategies. To prove our concept, we conducted Monte Carlo experiments for `no sharing', `active sharing', and `passive sharing', and compared the percolation probability and SINR coverage proportion in different sharing strategies, among which we showed that active sharing can provide the best coverage performance.  In addition, we provided different suggestions for the MNOs with different sizes of BS deployment.

\appendices
\section{Proof of Theorem \ref{theo:SCAandCA}}\label{app:SCAandCA}
To prove that the face percolation of CAs of BSs is the same as the face percolation of SCAs of BSs, we need to prove that:
\begin{equation}
\bigcup\limits_{i}\Psi_i\cap\Omega_i=\bigcup\limits_{i}\Psi_i.
\end{equation}
Using the SINR $\beta(z)$, the CAs can be defined as:
\begin{equation}
\begin{array}{r@{}l}
    \Psi_i
    &\overset{\triangle}{=}\bigg\{z\in\mathbb{R}^2:\beta_i(z)\geq\beta\bigg\},
\end{array}
\end{equation}
and the Voronoi cells can be defined as:
\begin{equation}
\begin{array}{r@{}l}
    \Omega_i
    &\overset{\triangle}{=}\bigg\{z\in\mathbb{R}^2:\beta_i(z)\geq\beta_j(z),\forall j\neq i\bigg\}.
\end{array}
\end{equation}
 We assume that $\bigcup_i \Omega_i=\Omega$, where  $\Omega$ is the universal set. Since $\Psi_i\cap \Omega_i \subseteq \Psi_i$ for any $i$, we can obtain:
\begin{equation}
    \bigcup_i \Psi_i\cap \Omega_i \subseteq \bigcup_i \Psi_i.
\end{equation}
Under our assumption, if a typical user at $z$ is covered by BS $x_j$ but  the closest BS is located at $x_i$, $z\in \Omega_i\cap\Psi_j$, we have $\beta_j(z)\geq \beta$ and $\beta_i(z)\geq \beta_j(z)$. Thus,  we have $\beta_i(z)\geq\beta$. Therefore, this UE is also covered by BS located at $x_i$, \ie $z\in\Psi_i$. Therefore, $\forall i\neq j$, $\Psi_j\cap \Omega_i\subseteq\Psi_i\cap\Omega_i$, and we have
\begin{equation}
\begin{array}{r@{}l}
    \displaystyle(\bigcup_j \Psi_j)\cap \Omega_i&\displaystyle=(\Psi_i\cap\Omega_i)\cup(\bigcup_{j\neq i}\Psi_j\cup\Omega_i)\\
    &=\Psi_i\cap\Omega_i,
\end{array}
\end{equation}
which means that the SCA $\Psi_i\cap\Omega_i$ contains all coverage areas inside the Voronoi cell $\Omega_i$. Next, we can also obtain:
\begin{equation}
\begin{array}{r@{}l}
    \displaystyle\bigcup_{i}\Psi_i\cap\Omega_i&\displaystyle=\bigcup_{i}\bigg((\bigcup_j \Psi_j)\cap \Omega_i\bigg)\\
    &\displaystyle=(\bigcup_j \Psi_j)\cap(\bigcup_{i}\Omega_i)\\
    &\displaystyle=(\bigcup_j \Psi_j)\cap\Omega\displaystyle=\bigcup_j \Psi_j,
\end{array}
\end{equation}
that is, the union of SCAs is the same as the union CAs.\\ 
\indent For CAs $\Psi_i$ and $\Psi_j$, if there exists $z\in\Psi_i\cap\Psi_j$, they are connected. But for SCAs $\Psi_i\cap\Omega_i$ and $\Psi_j\cap\Omega_j$, all the points on their intersection are located on their Voronoi boundary. Therefore, if there exist some points on the common boundary where the SINR $\beta_i(z)$ and $\beta_j(z)$ are larger or equal to the threshold $\beta$, $\Psi_i\cap\Omega_i$ and $\Psi_j\cap\Omega_j$ are connected.

\section{Proof of Theorem~\ref{theo:betagamma}}\label{app:betagamma}

We focus on a typical user $z$ on the common Voronoi boundary of two neighbour BSs located at $x_i$ and $x_j$. We assume that the distance between the user and these two BSs are both $d$, \ie $\|x_i-z\|=\|x_j-z\|=d$. The BSs at $x_i$ and $x_j$ provide the strongest received power for the typical user. Therefore, the SINR at $z$ satisfies:
\begin{equation}
\begin{array}{r@{}l}
    \beta_i&(z)=\beta_j(z)\\
    &\displaystyle=\frac{P_t d^{-\alpha}}{N_0+\gamma P_t d^{-\alpha} + \gamma\sum\limits_{x_k\in\Phi\backslash \{x_i,x_j\}}P_t l(x_k-z)}\\
    &\displaystyle<\frac{P_t d^{-\alpha}}{\gamma P_t d^{-\alpha}}=\frac{1}{\gamma}.
\end{array}
\end{equation}

\indent If $\frac{1}{\gamma}\leq \beta$, \ie $\beta\gamma\geq 1$, any points on the Voronoi boundary can not achieve enough SINR. Thus, no user can realize handover between different BSs' services, and there is no percolation. Therefore, a necessary condition for the non-zero percolation probability of a single-layer GDM is $\beta\gamma<1$. 
\section{Proof of Theorem \ref{theo:devicedegree}}\label{app:devicedegree}
\indent In cellular networks, we name the closest BS as the serving BS, and name other BSs that can also provide enough SINR larger than $\beta$ as `potential serving BSs'. Denote $N$ as the number of potential serving BSs. Let $x_{(k)}$ denote the $k$th closest potential serving BS to the typical user at $z$ and $x_{(0)}$ denote the location of the serving BS. Therefore, the SINR at $z$ from BS located at $x_{(N)}$ satisfies:
\begin{equation}
\begin{array}{r@{}l}
    \beta_{(N)}(z)&=\displaystyle
    \frac{P_t \|x_{(N)}-z\|^{-\alpha}}{N_0+\gamma\sum\limits_{j=0}^{N-1} P_t \|x_{(j)}-z\|^{-\alpha}}\\
    &< \displaystyle
    \frac{P_t \|x_{(N)}-z\|^{-\alpha}}{\gamma\sum\limits_{j=0}^{N-1} P_t \|x_{(j)}-z\|^{-\alpha}}\\
    &< \displaystyle
    \frac{P_t \|x_{(N)}-z\|^{-\alpha}}{\gamma N P_t \|x_{(N)}-z\|^{-\alpha}}=\frac{1}{\gamma N}.
\end{array}
\end{equation}

Because the $x_{(N)}$ is the farthest potential serving BS, its SINR at $z$ should be larger or equal to beta, \ie $\beta_{(N)}(z)\geq\beta$. Therefore, $\frac{1}{\gamma N}>\beta$, that is, $N<\frac{1}{\gamma \beta}$.
When $\beta\gamma\geq 1$, $N=0$ and there is no potential serving BSs and all users can be only served by the closest BS. In this case, percolation can not be realized because handovers between different BSs' coverage areas are impossible.

\section{Proof of Theorem~\ref{theo:lambdac1}}\label{app:lambdac1}

 \indent To analyze the continuity of coverage areas in cellular networks, we focus on the Gilbert disk model $D(\lambda,r)$ where $\lambda$ is the density of BSs and $r$ is defined as the coverage radius. The sufficient and necessary condition for phase transition of percolation probability is:
 \begin{equation}
     \lambda>\frac{\lambda_c(1)}{4r^2}.
     \label{conditionGD}
 \end{equation}
 \begin{figure}
    \centering
    \includegraphics[width=0.5\linewidth]{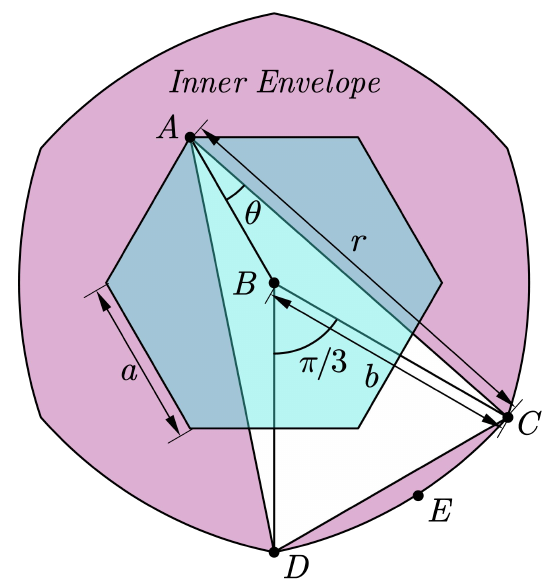}
    \caption{The `inner envelope' in Theorem \ref{theo:lambdac1}.}
    \label{fig:InnerEnvelope}
\end{figure}

 \indent But the value of 
$\lambda_c(1)$ is an open question. Let us focus on the first `special case'. If there is at least one BS inside the `inner envelope' shown in Fig.\ref{fig:InnerEnvelope} ($r\gg 2a$), the entire hexagon $\ncalH$ is completely covered. We have
\begin{equation}
    \P\{\ncalH {\rm \;is\; completely\; covered}\}=1-e^{-\lambda S_{\rm in}},
\end{equation}
where
\begin{equation}
    S_{\rm in}=6(\theta r^2-2S_{\bigtriangleup ABC}).
\end{equation}
is the area inside the inner envelope, $\theta$ is the radian of $\angle BAC$, and $S_{\bigtriangleup ABC}$ is the area of the triangle $\bigtriangleup ABC$. We define that $b=\|\overline{BC}\|$. Using the law of cosines, we have
\begin{equation}
\cos \frac{5\pi}{6}=\frac{a^2+b^2-r^2}{2ab}=-\frac{\sqrt{3}}{2}, 
\end{equation}
and we have $b=\sqrt{r^2-\frac{a^2}{4}}-\frac{\sqrt{3}}{2}a$. Using the law of sines, we have
\begin{equation}
    \frac{\sin \theta}{b}=\frac{\sin \frac{5\pi}{6}}{r},
\end{equation}
and $\theta=\arcsin{\frac{b}{2r}}$. The area of the triangle $\bigtriangleup ABC$ is 
\begin{equation}
    S_{\bigtriangleup ABC}=\frac{1}{2}ab\sin \frac{5\pi}{6}=\frac{1}{4}ab.
\end{equation}
\indent Therefore, the area of the inner envelope is 
\begin{equation}
    S_{\rm in}=6\theta r^2-3ab,
\end{equation}
which increases as $a$ decreases ($0<a<\frac{r}{2}$). When $a$ approaches 0, $S_{\rm in}$ approaches its maximum value $\pi r^2$.\\
\indent To achieve percolation of covered hexagons, the sufficient condition is:
\begin{equation}
    \P\{\ncalH {\rm \;is\; completely\; covered}\}=1-e^{-\lambda S_{\rm in}}>\frac{1}{2},
\end{equation}
which is equivalent to 
\begin{equation} \label{Sin}
    \lambda S_{\rm in}> \ln 2.
\end{equation}
\indent When the side length $a$ of hexagons approaches 0, all hexagons are much smaller than the coverage areas so that they can be considered discrete points. At the same time, (\ref{Sin}) becomes $\lambda \pi r^2 >\ln 2$. Notice that $\lambda \pi r^2 >\ln 2$ is the necessary but not sufficient condition for $\lambda S_{\rm in}> \ln 2$. Only when $a$ approaches 0, they are equivalent.

\begin{figure}
    \centering
    \includegraphics[width=0.8\linewidth]{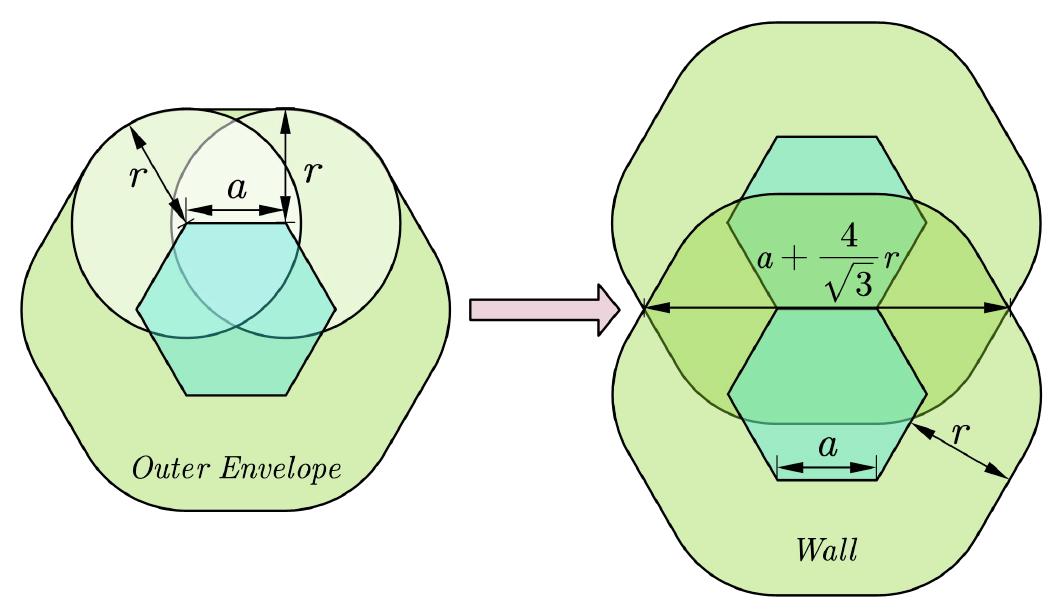}
    \caption{The `outer envelope' and `wall' in Theorem \ref{theo:lambdac1}.}
    \label{fig:OuterWall}
\end{figure}

\begin{figure}
    \centering
    \includegraphics[width=0.85\linewidth]{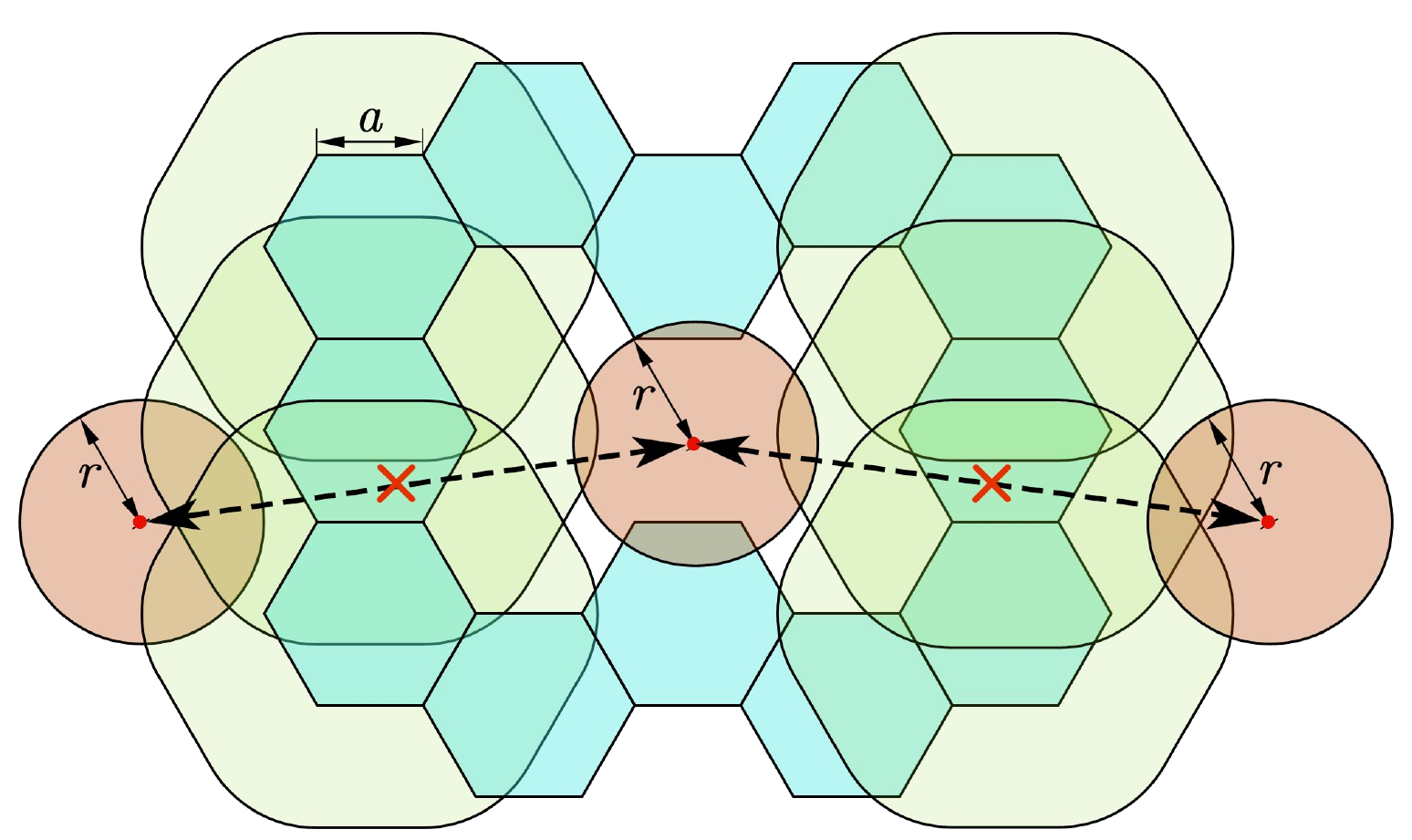}
    \caption{A closed circuit formed by many uncovered hexagons. There are many adjacent uncovered hexagons and `walls' to prevent coverage areas inside or outside the closed circuit from being connected.}
    \label{fig:ClosedCircuit}
\end{figure}
\indent Next, let us focus on another `special case'. If there is no BS inside the `outer envelope' shown in Fig.\ref{fig:OuterWall}, the entire hexagon $\ncalH$ is completely not covered. We have
\begin{equation}
    \P\{\ncalH {\rm \;is\; completely\; not\; covered}\}=e^{-\lambda S_{\rm out}}>\frac{1}{2},
\end{equation}
where 
\begin{equation}
    S_{\rm out}=\frac{3\sqrt{3}}{2}a^2+\pi r^2+6ar.
\end{equation}
\indent As $a$ decreases, the area of the outer envelope $S_{\rm out}$ decreases. When $a$ approaches 0, $S_{\rm out}$ approaches its minimum value $\pi r^2$. In order to achieve percolation of uncovered hexagons, the sufficient condition is:
\begin{equation}
    \P\{\ncalH {\rm \;is\; completely\;not \; covered}\}=e^{-\lambda S_{\rm out}}>\frac{1}{2},
\end{equation}
which is equivalent to 
\begin{equation} \label{Sout}
    \lambda S_{\rm out}<\ln 2.
\end{equation}
\indent Similarly, decreasing the value of side length $a$, all hexagons are much smaller than coverage areas of BSs and (\ref{Sout}) becomes $\lambda \pi r^2 <\ln 2$. Notice that $\lambda \pi r^2 <\ln 2$ is also only the necessary but not sufficient condition for $\lambda S_{\rm out}< \ln 2$. They are equivalent when $a$ approaches 0 (much less than $r$).\\
\indent Only the percolation of uncovered hexagons is not enough to prove that the percolation probability of coverage areas is $0$. As shown in Fig.\ref{fig:OuterWall}, two adjacent uncovered hexagons form a `wall' to avoid the continuity of coverage areas. The minimum distance between BSs on different sides of the `wall' is $a+\frac{4}{\sqrt{3}}r$ which is always larger than $2r$. That means the wall formed by two adjacent uncovered hexagons prevents the connection of coverage areas of BSs on both sides of this wall. We can assume that there is a BS in the origin. When $\P\{\ncalH {\rm \;is\; completely\; not\; covered}\}=e^{-\lambda S_{\rm out}}>\frac{1}{2}$, a circuit of uncovered hexagons (as shown in Fig.\ref{fig:ClosedCircuit}) is formed around the origin, which consists of many walls that can prevent the continuity of coverage areas.\\
\indent For the Gilbert disk model, the transformation of scaling does not affect the percolation state of a random graph and the percolation probability of $D(\lambda,r)$. Mathematically, $\forall A>0$, $D(\lambda,r)$ and $D(A^2\lambda,r/A)$ have the same percolation probability. This indicates that the critical value of $\lambda \pi r^2$, the product of density $\lambda$ and coverage area $\pi r^2$, is a vital factor. Such a concept is also mentioned in \cite{haenggi2012stochastic}, which means the continuous percolation can be analyzed using discrete percolation. When the side length of considered hexagons is much less than the coverage radius, the critical value of $\lambda\pi r^2$ is $\ln 2$. Because we have defined $\lambda(1)=\lambda (2r)^2$, its critical value is $\lambda_c(1)=4\ln 2/\pi$. 

\section{Proof of Theorem~\ref{theo:acr}}\label{app:acr}
We consider a single-layer cellular network where all BSs interfere with each other. Assume that $\|x_{i}-z\|$ is the distance from a typical UE at $z$ to a considered BS located at $x_{i}$. We define $r_{1}=\|x_{1}-z\|=\min\{\|x_{i}-z\|\}$, where $x_{1}$ is the location of the nearest BS to typical UE and we assume that $r_{1}>1\,{\rm m}$. Based on (\ref{betaiz}), the SINR from the considered BS to the typical UE is 
\begin{equation}
    \beta_{i}(z)=\frac{P_t \|x_{i}-z\|^{-\alpha}}{N_0+\gamma\sum\limits_{x_{j}\in\Phi\setminus x_{i}}P_t\|x_{j}-z\|^{-\alpha}}.
\end{equation}
\indent Therefore, the typical UE at $y$ can be covered by cellular networks when $\max\{\beta_{i}(z)\}\geq\beta$, where $\beta$ is the SINR threshold. This event can be also expressed as:
\begin{equation}
\frac{P_t r_{1}^{-\alpha}}{N_0+\gamma P_t\sum\limits_{x_{j}\in\Phi\setminus x_{1}}\|x_{j}-z\|^{-\alpha}}\geq\beta.
\end{equation}

\begin{figure}
    \centering
    \includegraphics[width=0.9\linewidth]{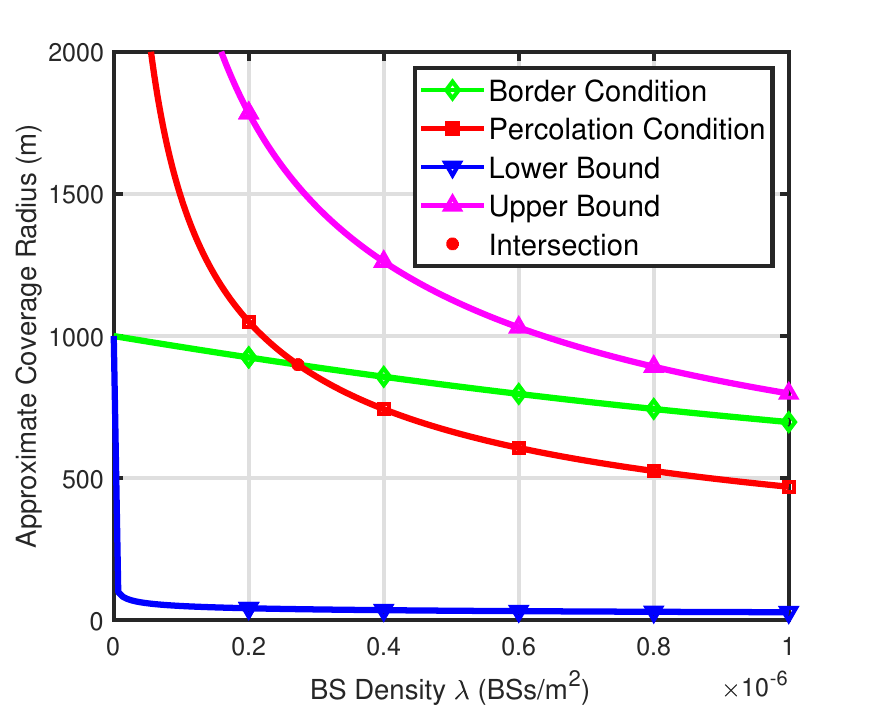}
    \caption{The implicit function $r_{m}(\lambda)$ introduced in (\ref{appro}) and (\ref{appro2}) is shown as the curve `Border Condition'. It is between the lower bound and upper bound that are shown in (\ref{inequality3}). For each density $\lambda$, the solution of approximate coverage radius is unique. Equation (\ref{percona}) is also shown as the `Percolation Condition' curve. There is only one intersection of the `Percolation Condition' and `Border Condition' curves.}
    \label{fig:onlysolution}
\end{figure}

Consider the users at the boundary of CAs, the distances to BSs satisfy:
\begin{equation}\label{nosharing1}
    r_{1}^{-\alpha}=\frac{\beta}{P_t/N_0}+\gamma\beta \sum\limits_{x_{j}\in\Phi\setminus x_{1}}\|x_{j}-z\|^{-\alpha}.
\end{equation}
Especially for the users on the boundary of SCAs, the distances to all interfering BSs are larger than $r_1$. \\
\indent In order to apply the Gilbert disk model to analyze the critical condition for percolation, we propose a parameter `average coverage radius' to approximate the BS's coverage range. The serving BS is at a distance $r_m$ to the typical user and the locations of interfering BSs follow a PPP with density $\lambda$ outside the circular area centered at the user with radius $r$. 
Using Campbell's theorem, we use an integral to calculate the expectation of interference conditioned on $r_m$:
\begin{equation}
\begin{array}{@{}r@{}l}
  \E\bigg[\sum\limits_{x_{j}\in\Phi\setminus x_{1}}\|x_{j}-z\|^{-\alpha}\bigg]&=\displaystyle\int_{r_{m}}^{\infty}r^{-\alpha}2\lambda\pi rdr\\
  &\displaystyle=\frac{2\pi \lambda}{\alpha-2}r_{m}^{2-\alpha}.
  \end{array}  
\end{equation}
\indent Therefore, $r_m$ is the solution of the following equation:
\begin{equation}
    r^{-\alpha}=\frac{\beta}{P_t/N_0}+\frac{2\pi \gamma\beta\lambda}{\alpha-2}r^{2-\alpha}.
\label{appro}
\end{equation}
Multiply $r^{\alpha}$ on both sides of (\ref{appro}), we have:
\begin{equation}\label{appro2}
    \frac{\beta}{P_t/N_0}r^{\alpha}+\frac{2\pi \gamma\beta\lambda}{\alpha-2}r^{2}=1.
\end{equation}
\indent Define that
\begin{equation}
\ncalF(r,\lambda)=\frac{\beta}{P_t/N_0}r^{\alpha}+\frac{2\pi \gamma\beta\lambda}{\alpha-2}r^{2},
\label{Frlam}
\end{equation}
which is an increasing function of $r$ when $\alpha>2$. When $r=0$,  $\ncalF(r,\lambda)=0$. When $r>(\frac{P_t/N_0}{\beta})^{\frac{1}{\alpha}}$ or $r>(\frac{\alpha-2}{2\pi\gamma\beta\lambda})^{\frac{1}{2}}$, $\ncalF(r,\lambda)>1$. Therefore, the solution of (\ref{appro2}), $r_m$, is unique which satisfies $0<r_{m}<\min\{(\frac{P_t/N_0}{\beta})^{\frac{1}{\alpha}},(\frac{\alpha-2}{2\pi\gamma\beta\lambda})^{\frac{1}{2}}\}$. For a higher $\lambda$, the value of $r_m$ becomes smaller. Therefore, the average coverage radius $r_{m}$ can be written as a non-increasing function of $\lambda$, and the `border condition' in Fig.\ref{fig:onlysolution} is the curve of the implicit function $r_{m}(\lambda)$ hiding in (\ref{appro2}). \\
\indent To approximate the actual network deployment as much as possible, we assume that all BSs can always cover the UEs inside their circular coverage areas with a unit coverage radius. Therefore, the average coverage radius $r_m$ should be larger than 1\;{\rm m}. Based on this, we have 
\begin{equation}
\ncalF(1,\lambda)=\displaystyle\frac{\beta}{P_t/N_0}+\frac{2\pi \gamma\beta\lambda}{\alpha-2}<1.
\end{equation}

\indent From (\ref{Frlam}), because $r_m^{\alpha}>r_m^{2}$, we have
\begin{equation}
\displaystyle\mathcal{F}(1,\lambda)r_{m}^{2}<\ncalF(r_{m},\lambda)
<\displaystyle\mathcal{F}(1,\lambda)r_{m}^{\alpha}.
\label{inequality2}
\end{equation}

Since $\mathcal{F}(r_m,\lambda)=1$,  $r_{m}$ satisfies 
\begin{equation}
\begin{array}{@{}r@{}l}
\displaystyle\mathcal{F}(1,\lambda)^{-\frac{1}{\alpha}}<r_{m}<\mathcal{F}(1,\lambda)^{-\frac{1}{2}}.
\end{array}
\label{inequality3}
\end{equation}
\indent The inequality (\ref{inequality3}) is also shown in Fig.\ref{fig:onlysolution}. The curve of $r_{m}(\lambda)$, \ie `border condition', is between the `lower bound' $r_L(\lambda)=\mathcal{F}(1,\lambda)^{-\frac{1}{\alpha}}$ and the `upper bound' $r_U(\lambda)=\mathcal{F}(1,\lambda)^{-\frac{1}{2}}$.\\

\section{Proof of Theorem \ref{theo:percov}}\label{app:pcov}
\indent To analyze continuous percolation using discrete percolation in hexagons, we focus on the `inner envelope' and `outer envelope' in Fig.\ref{fig:InnerEnvelope} and Fig.\ref{fig:OuterWall}. The areas of them satisfy:
\begin{equation}
    S_{\rm in}<\pi r^2<S_{\rm out}.
\end{equation}
\indent When the side length $a$ of considered hexagons is much less than the coverage radius, the shapes of `inner envelope' and `outer envelope' are almost the same as a circular area with radius $r$ and the areas of them both converge to $\pi r^2$, \ie
\begin{equation}
    \lim\limits_{a\ll r, a\rightarrow 0^+}S_{\rm in}=\lim\limits_{a\ll r, a\rightarrow 0^+}S_{\rm out}=\pi r^2.
\end{equation}

When $a$ approaches 0, the discrete percolation of hexagons is used to approximate the continuous percolation of points. The probability of a point being covered (\ie coverage probability) is 
\begin{equation}
    p_{cov}=\lim\limits_{a\rightarrow 0^+}1-e^{-\lambda S_{\rm in}}=1-e^{-\lambda\pi r^2},
\end{equation} 
and the probability of a point being not covered is
\begin{equation}
    \lim\limits_{a\rightarrow 0^+}e^{-\lambda S_{\rm out}}=e^{-\lambda\pi r^2}=1-p_{cov}.
\end{equation} 
\indent Similar to Theorem \ref{theo:lambdac1}, if $p_{cov}$ is larger than $1/2$, the percolation probability is non-zero. If $1-p_{cov}$ is larger than $1/2$, the percolation probability is zero, while $p_{cov}$ is less than $1/2$ at the same time. Therefore, for continuous percolation, the phase of percolation probability (zero or non-zero) depends on whether $p_{cov}$ is greater than $1/2$ or not.

\section{Proof of Lemma~\ref{lem:criconlam}}\label{app:criconlam}
\indent In Theorem \ref{theo:acr}, we obtain the relationship between BS density $\lambda_a$ and its corresponding average coverage radius $r_{a}$. Based on (\ref{GDMcricon}), the critical condition for phase transition of percolation probability is 
\begin{equation}\label{GDMcricona}
\lambda_a r_{a}^2=\frac{\lambda_c(1)}{4}.
\end{equation}
\indent Let $\beta_0=P_t/N_0$, $r_a=r_{m}(\lambda_a)$ and substitute (\ref{GDMcricona}) into (\ref{appro2}), we have:
\begin{equation}\label{percona}
    \frac{\beta}{\beta_0}r_{a}^{\alpha}+\frac{\pi\gamma\beta\lambda_c(1)}{2(\alpha-2)}=1.
\end{equation}

Therefore, the average coverage radius can be derived as:
\begin{equation}\label{crirad}
    r_{a}=\bigg(\frac{\beta_0}{\beta}\bigg(1-\frac{\pi \gamma \beta\lambda_c(1)}{2(\alpha-2)}\bigg)\bigg)^{\frac{1}{\alpha}}.
\end{equation}
\indent Substitute (\ref{crirad}) into (\ref{GDMcricona}), the critical value of BS density is 
\begin{equation}\label{lambdaac}
    \lambda_a=\frac{\lambda_c(1)}{4}\bigg(\frac{\beta_0}{\beta}\bigg(1-\frac{\pi \gamma \beta\lambda_c(1)}{2(\alpha-2)}\bigg)\bigg)^{-\frac{2}{\alpha}}.
\end{equation}
\indent The `percolation condition' in Fig.\ref{fig:onlysolution} shows the critical relationship in (\ref{GDMcricona}). There is only one intersection between the `border condition' and `percolation condition', which shows the unique solution of critical BS density in (\ref{lambdaac}).  


\section{Proof of Lemma~\ref{lem:2GDpre}}\label{app:2GDpre}
\indent In the graph $G(V,E)$, the sets of BSs' locations of MNO $a$ and MNO $b$ are modeled using independent PPPs $\Phi_a$ with density $\lambda_a$ and $\Phi_b$ with density $\lambda_b$. The vertice set $V$ is the superposition of $\Phi_a$ and $\Phi_b$, \ie $V=\Phi_a \cup \Phi_b$, which can be modeled as a PPP with density $\lambda_a+\lambda_b$. Approximated by the superposition of two Gilbert disk models $D(\lambda_a,r_{a})$ and $D(\lambda_b,r_{b})$, the edge set $E$ can be approximated using $E_a\cup E_b \cup E_{ab}$. The edge set $E_a$ is
\begin{equation}
    E_a=\{\overline{x_{i}x_{j}}:\|x_{i}-x_{j}\|\leq 2r_{a},\,\forall\,x_{i},x_{j}\in\Phi_a\},
\end{equation}
which represents the connectivity between coverage areas of MNO $a$'s BSs. Similarly, the edge set $E_b$ represents the connectivity between coverage areas of MNO $b$'s BSs:
\begin{equation}
    E_b=\{\overline{x_{i}x_{j}}:\|x_{i}-x_{j}\|\leq 2r_{b},\,\forall\,x_{i},x_{j}\in\Phi_b\}.
\end{equation}
\indent Differently, the set $E_{ab}$ represents the connectivity between MNO $a$'s coverage areas and MNO $b$'s coverage areas, which is expressed as:
\begin{equation}
\begin{array}{@{}r@{}l}
    E_{ab}=\{\overline{x_{i}x_{j}}:&\|x_{i}-x_{j}\|\leq r_{a}+r_{b},\\ &\;\;\;\;\;\;\;\;\;\;\;\;\;\;\;\;\;\;\;\;\forall\,x_{i}\in\Phi_a,x_{j}\in\Phi_b\}.  
\end{array}
\end{equation}
\indent Next, we prove the sufficient condition for zero and non-zero percolation probability. We first consider an extreme graph $G(V,E_L)=\{V,E_L\}$, where the edge set $E_L$ is
\begin{equation}
\begin{array}{@{}r@{}l}
    E_L=\{\overline{x_{i}x_{j}}:\|x_{i}-x_{j}\|\leq & 2\max\{r_{a},r_{b}\},\\
    &\;\;\;\;\;\;\;\;\;\;\;\forall\,x_{i},x_{j}\in V\}.
\end{array}
\end{equation}
\indent At the same time, consider another extreme graph $G(V,E_U)=\{V,E_U\}$, where the edge set $E_U$ is
\begin{equation}
\begin{array}{@{}r@{}l}
    E_U=\{\overline{x_{i}x_{j}}:\|x_{i}-x_{j}\|\leq &2\min\{r_{a},r_{b}\},\\
    &\;\;\;\;\;\;\;\;\;\;\;\forall\,x_{i},x_{j}\in V\}.
\end{array}
\end{equation}
\indent We can obtain the relationship between the edge sets:
\begin{equation}
    E_U\subseteq E,\, E\subseteq E_L.
\end{equation}
\indent Therefore, the relationship between $G(V,E)$, $G(V,E_L)$ and $G(V,E_U)$ is shown as follows:
\begin{equation}
    G(V,E_U)\subseteq G(V,E),\,G(V,E)\subseteq G(V,E_L).
    \label{relationship}
\end{equation}
\indent As a common understanding, the critical value of BS density $\lambda$ for a GDM $D(\lambda,r)$ is $\lambda_c(1)/4r^2$. Therefore, the sufficient condition for $\theta(\lambda_a+\lambda_b,\max\{r_{a},r_{b}\})=0$ is 
\begin{equation}
    \lambda_a+\lambda_b<\frac{\lambda_c(1)}{4\max\{r_{a},r_{b}\}^2}.
\label{lower}
\end{equation}
\indent Similarly, we can also obtain the sufficient condition for $\theta(\lambda_a+\lambda_b,\min\{r_{a},r_{b}\})>0$: 
\begin{equation}
    \lambda_a+\lambda_b>\frac{\lambda_c(1)}{4\min\{r_{a},r_{b}\}^2}.
\label{upper}
\end{equation}
\indent According to the relationship between $G(V,E)$, $G(V,E_L)$ and $G(V,E_U)$ shown in (\ref{relationship}), (\ref{lower}) is the sufficient condition for $\theta(\lambda_a,\lambda_b)=0$ and (\ref{upper}) is the sufficient condition for $\theta(\lambda_a,\lambda_b)>0$. These two conditions form the restriction region of the critical condition for phase transition of percolation probability. 
\section{Proof of Lemma~\ref{lem:2GD}} \label{app:2GD}

\indent For active sharing case, we want to analyze the critical condition for percolation of coverage areas when we have two independent Gilbert disk models $D(\lambda_a,r_{a})$ and $D(\lambda_b,r_{b})$. In Theorem \ref{theo:percov}, we prove that the percolation probability is related to the coverage probability of any point. In this case, the probability of a point being covered is $p_{cov}=1-e^{-\lambda_a \pi r_{a}^2 -\lambda_b \pi r_{b}^2}$ and the probability of a point being not covered is $1-p_{cov}=e^{-\lambda_a \pi r_{a}^2 -\lambda_b \pi r_{b}^2}$. Based on (\ref{percov}) in Theorem \ref{theo:percov}, the critical condition for the phase transition of percolation probability can be simplified as: 
\begin{equation}
    \lambda_a \pi r_a^2 + \lambda_b \pi r_b^2 = \ln 2.
\end{equation}

\ifCLASSOPTIONcaptionsoff
  \newpage
\fi

\bibliographystyle{IEEEtran}
\input{main.bbl}



\end{document}

%% file: notation.tex
\def\nba{{\mathbf{a}}}
\def\nbb{{\mathbf{b}}}
\def\nbc{{\mathbf{c}}}
\def\nbd{{\mathbf{d}}}
\def\nbe{{\mathbf{e}}}
\def\nbf{{\mathbf{f}}}
\def\nbg{{\mathbf{g}}}
\def\nbh{{\mathbf{h}}}
\def\nbi{{\mathbf{i}}}
\def\nbj{{\mathbf{j}}}
\def\nbk{{\mathbf{k}}}
\def\nbl{{\mathbf{l}}}
\def\nbm{{\mathbf{m}}}
\def\nbn{{\mathbf{n}}}
\def\nbo{{\mathbf{o}}}
\def\nbp{{\mathbf{p}}}
\def\nbq{{\mathbf{q}}}
\def\nbr{{\mathbf{r}}}
\def\nbs{{\mathbf{s}}}
\def\nbt{{\mathbf{t}}}
\def\nbu{{\mathbf{u}}}
\def\nbv{{\mathbf{v}}}
\def\nbw{{\mathbf{w}}}
\def\nbx{{\mathbf{x}}}
\def\nby{{\mathbf{y}}}
\def\nbz{{\mathbf{z}}}
\def\nb0{{\mathbf{0}}}
\def\nb1{{\mathbf{1}}}

\def\nbA{{\mathbf{A}}}
\def\nbB{{\mathbf{B}}}
\def\nbC{{\mathbf{C}}}
\def\nbD{{\mathbf{D}}}
\def\nbE{{\mathbf{E}}}
\def\nbF{{\mathbf{F}}}
\def\nbG{{\mathbf{G}}}
\def\nbH{{\mathbf{H}}}
\def\nbI{{\mathbf{I}}}
\def\nbJ{{\mathbf{J}}}
\def\nbK{{\mathbf{K}}}
\def\nbL{{\mathbf{L}}}
\def\nbM{{\mathbf{M}}}
\def\nbN{{\mathbf{N}}}
\def\nbO{{\mathbf{O}}}
\def\nbP{{\mathbf{P}}}
\def\nbQ{{\mathbf{Q}}}
\def\nbR{{\mathbf{R}}}
\def\nbS{{\mathbf{S}}}
\def\nbT{{\mathbf{T}}}
\def\nbU{{\mathbf{U}}}
\def\nbV{{\mathbf{V}}}
\def\nbW{{\mathbf{W}}}
\def\nbX{{\mathbf{X}}}
\def\nbY{{\mathbf{Y}}}
\def\nbZ{{\mathbf{Z}}}

\def\ncalA{{\mathcal{A}}}
\def\ncalB{{\mathcal{B}}}
\def\ncalC{{\mathcal{C}}}
\def\ncalD{{\mathcal{D}}}
\def\ncalE{{\mathcal{E}}}
\def\ncalF{{\mathcal{F}}}
\def\ncalG{{\mathcal{G}}}
\def\ncalH{{\mathcal{H}}}
\def\ncalI{{\mathcal{I}}}
\def\ncalJ{{\mathcal{J}}}
\def\ncalK{{\mathcal{K}}}
\def\ncalL{{\mathcal{L}}}
\def\ncalM{{\mathcal{M}}}
\def\ncalN{{\mathcal{N}}}
\def\ncalO{{\mathcal{O}}}
\def\ncalP{{\mathcal{P}}}
\def\ncalQ{{\mathcal{Q}}}
\def\ncalR{{\mathcal{R}}}
\def\ncalS{{\mathcal{S}}}
\def\ncalT{{\mathcal{T}}}
\def\ncalU{{\mathcal{U}}}
\def\ncalV{{\mathcal{V}}}
\def\ncalW{{\mathcal{W}}}
\def\ncalX{{\mathcal{X}}}
\def\ncalY{{\mathcal{Y}}}
\def\ncalZ{{\mathcal{Z}}}

\def\nbbA{{\mathbb{A}}}
\def\nbbB{{\mathbb{B}}}
\def\nbbC{{\mathbb{C}}}
\def\nbbD{{\mathbb{D}}}
\def\nbbE{{\mathbb{E}}}
\def\nbbF{{\mathbb{F}}}
\def\nbbG{{\mathbb{G}}}
\def\nbbH{{\mathbb{H}}}
\def\nbbI{{\mathbb{I}}}
\def\nbbJ{{\mathbb{J}}}
\def\nbbK{{\mathbb{K}}}
\def\nbbL{{\mathbb{L}}}
\def\nbbM{{\mathbb{M}}}
\def\nbbN{{\mathbb{N}}}
\def\nbbO{{\mathbb{O}}}
\def\nbbP{{\mathbb{P}}}
\def\nbbQ{{\mathbb{Q}}}
\def\nbbR{{\mathbb{R}}}
\def\nbbS{{\mathbb{S}}}
\def\nbbT{{\mathbb{T}}}
\def\nbbU{{\mathbb{U}}}
\def\nbbV{{\mathbb{V}}}
\def\nbbW{{\mathbb{W}}}
\def\nbbX{{\mathbb{X}}}
\def\nbbY{{\mathbb{Y}}}
\def\nbbZ{{\mathbb{Z}}}

\def\nfrakR{{\mathfrak{R}}}

\def\nrma{{\rm a}}
\def\nrmb{{\rm b}}
\def\nrmc{{\rm c}}
\def\nrmd{{\rm d}}
\def\nrme{{\rm e}}
\def\nrmf{{\rm f}}
\def\nrmg{{\rm g}}
\def\nrmh{{\rm h}}
\def\nrmi{{\rm i}}
\def\nrmj{{\rm j}}
\def\nrmk{{\rm k}}
\def\nrml{{\rm l}}
\def\nrmm{{\rm m}}
\def\nrmn{{\rm n}}
\def\nrmo{{\rm o}}
\def\nrmp{{\rm p}}
\def\nrmq{{\rm q}}
\def\nrmr{{\rm r}}
\def\nrms{{\rm s}}
\def\nrmt{{\rm t}}
\def\nrmu{{\rm u}}
\def\nrmv{{\rm v}}
\def\nrmw{{\rm w}}
\def\nrmx{{\rm x}}
\def\nrmy{{\rm y}}
\def\nrmz{{\rm z}}

\def\nbydef{:=}
\def\nborel{\ncalB(\nbbR)}
\def\nboreld{\ncalB(\nbbR^d)}
\def\sinc{{\rm sinc}}

\newtheorem{lemma}{Lemma}
\newtheorem{thm}{Theorem}
\newtheorem{definition}{Definition}
\newtheorem{ndef}{Definition}
\newtheorem{nrem}{Remark}
\newtheorem{theorem}{Theorem}
\newtheorem{prop}{Proposition}
\newtheorem{cor}{Corollary}
\newtheorem{example}{Example}
\newtheorem{remark}{Remark}
\newtheorem{assumption}{Assumption}
	

\newcommand{\ceil}[1]{\lceil #1\rceil}
\def\argmin{\operatorname{arg~min}}
\def\argmax{\operatorname{arg~max}}
\def\figref#1{Fig.\,\ref{#1}}%
\def\E{\mathbb{E}}
\def\EE{\mathbb{E}^{!o}}
\def\P{\mathbb{P}}
\def\pc{\mathtt{P_c}}
\def\rc{\mathtt{R_c}}   
\def\p{p}

\def\V{\operatorname{Var}}
\def\erfc{\operatorname{erfc}}
\def\erf{\operatorname{erf}}
\def\opt{\mathrm{opt}}
\def\R{\mathbb{R}}
\def\Z{\mathbb{Z}}

\def\LL{\mathcal{L}^{!o}}
\def\var{\operatorname{var}}
\def\supp{\operatorname{supp}}

\def\N{\sigma^2}
\def\T{\beta}							
\def\sinr{\mathtt{SINR}}			
\def\snr{\mathtt{SNR}}
\def\sir{\mathtt{SIR}}
\def\ase{\mathtt{ASE}}
\def\se{\mathtt{SE}}

\def\calN{\mathcal{N}}
\def\FE{\mathcal{F}}
\def\calA{\mathcal{A}}
\def\calK{\mathcal{K}}
\def\calT{\mathcal{T}}
\def\calB{\mathcal{B}}
\def\calE{\mathcal{E}}
\def\calP{\mathcal{P}}
\def\calL{\mathcal{L}}


\def\l{\ell}
\newcommand{\fad}[2]{\ensuremath{\mathtt{h}_{#1}[#2]}}
\newcommand{\h}[1]{\ensuremath{\mathtt{h}_{#1}}}

\newcommand{\err}[1]{\ensuremath{\operatorname{Err}(\eta,#1)}}
\newcommand{\FD}[1]{\ensuremath{|\mathcal{F}_{#1}|}}



\def\Bx{{\mathcal{B}}^x}
\def\Bxx{{\mathcal{B}}^{x_0}}
\def\jx{y}
\def\m{(\bar{n}-1)}
\def\mm{\bar{n}-1}
\def\Nx{{\mathcal{N}}^x}
\def\Nxo{{\mathcal{N}}^{x_0}}
\def\wj{w_{jx_0}}
\def\uij{u_{jx}}
 \def\yj{y}
 \def\yjx{y}
 \def\zjx{z_x}
 \def \tx {y_0}
 \def \htx {h_0}

\def\rx{z_{1}}
\def\ry{z_{2}}

\def\Rx{Z_{1}}
\def\Ry{Z_{2}}

\def \hyxx {h_{y_{x_0}}}
\def \hyx {h_{y_x}}

\def\nbb1{\mathbbm{1}}
\def\xi{x_i}
\def\xj{x_j}
\def\xx{x_0}
\def\yk{y_k}
\def\yy{y_0}
\def\ie{{\em i.e. }}
\def\eg{{\em e.g. }}
\def\iid{{\em i.i.d. }}
\def\avg{\rm avg}

\def\rmnuma{\rm\uppercase\expandafter{\romannumeral1}}
\def\rmnumb{\rm\uppercase\expandafter{\romannumeral2}}
\def\rmnumc{\rm\uppercase\expandafter{\romannumeral3}}
\def\rmnumd{\rm\uppercase\expandafter{\romannumeral4}}
\def\rmnume{\rm\uppercase\expandafter{\romannumeral5}}
\def\rmnumf{\rm\uppercase\expandafter{\romannumeral6}}